%% file: quant-hscc.tex
\documentclass{sig-alternate}

\usepackage{url}

\usepackage{algorithm2e}

\usepackage{graphicx}
\usepackage{amssymb}
\usepackage{color}
\usepackage{graphics}
\usepackage{epsfig}
\usepackage{latexsym}
\usepackage{amsmath}
\usepackage{xspace}

\usepackage{graphicx}
\usepackage{subfigure}

\input{macros.tex}

\newdef{definition}{Definition}
\newtheorem{theorem}{Theorem}
\newtheorem{lemma}{Lemma}

\newtheorem{corollary}{Corollary}

\begin{document}

\title{Synthesizing Switching Logic to Minimize Long-Run Cost }
\numberofauthors{3}
\author{
\alignauthor
Susmit Jha \\
\affaddr{UC Berkeley}\\
\email{jha@eecs.berkeley.edu}
\alignauthor
Sanjit A. Seshia \\
\affaddr{UC Berkeley}\\
\email{sseshia@eecs.berkeley.edu}
\alignauthor
Ashish Tiwari \\
\affaddr{SRI International}\\
\email{tiwari@csl.sri.com}
}

\maketitle
\begin{abstract}
Given a multi-modal dynamical system, optimal switching logic synthesis involves generating the conditions for switching between the system modes such that the resulting hybrid system satisfies a quantitative specification. We formalize and solve the problem of optimal switching logic synthesis for quantitative specifications over long run behavior. Our paper generalizes earlier work on synthesis for safety.  We present an approach for specifying quantitative measures using reward and penalty functions, and illustrate its effectiveness using several examples. Each trajectory of the system, and each state of the system, is associated with a cost. Our goal is to synthesize a system that minimizes this cost from each initial state. We present an automated technique to synthesize switching logic for such quantitative measures. Our algorithm works in two steps. For a single initial state, we reduce the synthesis problem to an unconstrained numerical optimization problem which can be solved by any off-the-shelf numerical optimization engines. In the next step, optimal switching condition is learnt as a generalization of the optimal switching states discovered for each initial state. We prove the correctness of our technique and demonstrate the effectiveness of this approach with experimental results.
\end{abstract}


\section{Introduction}
\label{sec:intro}
\input{intro.tex}

\section{Problem Definition}
\label{sec:prob}
\input{prob.tex}

\input{related.tex}

\section{Optimization Formulation}
\label{sec-formulate}
\input{optimize.tex}

\section{Optimization Algorithm}
\label{sec-algo}
\input{algo.tex}

\section{Multiple Initial States}
\label{sec-multiinit}
\input{multi.tex}

\section{Case Studies}
\label{sec-cases}
\input{exp.tex}

\section{Conclusion}
In this paper, we present an algorithm for automated
synthesis of switching logic in order to achieve
minimum long-run cost. Our algorithm is based
on reducing the switching logic synthesis problem
to an unconstrained numerical optimization problem
which can then be solved by existing optimization
techniques. We also give a learning-based approach
to generalize from a sample of switching states to
a switching condition, where the learnt condition is
optimal with high probability.

\bibliographystyle{abbrv}   
\bibliography{paper}       

\noop{
\appendix

\section{Quantiative Cost Examples}
\label{app-costEx}
\input{costEx.tex}

\section{Limit Behaviors}
\label{app-limitbehave}
\input{limitbehave.tex}

\section{Constrained Optimization Example}
\label{app-toyopt}
\input{toyopt.tex}

\section{Proof of Lemma 1}
\label{app-lemmaproof}
\input{lemmadecompose.tex}

\section{Proof of Theorem 2}
\label{app-multiproof} 
\input{multiproof.tex}


\section{Circuit of DC-DC Converter}
The circuit diagram for the DC-DC converter is 
presented in Figure~\ref{fig:boostcd}.
The parameters used in the experiments are as follows:
$V_d = 5V, C = 3.3 \mu F, L = 47 \mu H, rC = 0.06\; \text{ohms},
rL = 0.1\; \text{ohms}, 
rd = 0.05\;\text{ohms},
rs = 0.05\;\text{ohms}
$.
Input $E$ applied is $10$ volts and the target voltage at
the load $R$ is $5$ volts. The load resistance periodically
varies between $R=100$ and $R=200$ at intervals of $0.6$ milliseconds.
\label{app:boost}
\begin{figure}[htpb]
\begin{center}
\includegraphics[width=3in] {boost.eps}
\caption{Circuit Diagram for DC-DC Boost Converter}
\label{fig:boostcd}
\end{center}
\end{figure}

}

%
%


\end{document}

%% file: macros.tex
\newcommand{\pumpon}{{\mathtt{ON}}}
\newcommand{\pumpoff}{{\mathtt{OFF}}}

\newcommand{\thermoHeat}{{\mathtt{HEAT}}}
\newcommand{\thermoCool}{{\mathtt{COOL}}}
\newcommand{\thermoOff}{{\mathtt{OFF}}}
\newcommand{\thermoTemp}{{\mathtt{temp}}}
\newcommand{\thermoOut}{{\mathtt{out}}}
\newcommand{\comfort}{{\mathtt{discomfort}}}
\newcommand{\fuel}{{\mathtt{fuel}}}
\newcommand{\weartear}{{\mathtt{swTear}}}
\newcommand{\thermotime}{{\mathtt{time}}}
\newcommand{\inff}{{\infty}}

\DeclareMathOperator*{\argmin}{arg\,min}

\newcommand {\noop}[1]{}
\newcommand {\extended}[1]{}

\newcommand {\forcav}[1]{}
\newcommand {\foremsoft}[1]{#1}

\newcommand{\period}{{L}}

\newcommand{\prtraj}{{\traj_{pref}}}
\newcommand{\reptraj}{{\traj_{rep}}}

\newcommand{\real}{{\mathbb{R}}}

\newcommand {\switchlogic} {\mathtt{SwL}} 
\newcommand{\init}{{\mathtt{Init}}}
\newcommand{\traj}{{\bf{qx}}}
\newcommand{\ptraj}{{\bf{qx_p}}}
\newcommand{\trajx}{{\bf{x}}}
\newcommand{\PR}{{\mathtt{PR}}}
\newcommand{\trajPR}{{\mathbf{PR}}}
\newcommand{\trajP}{{\mathbf{P}}}
\newcommand{\trajR}{{\mathbf{R}}}

\newcommand{\trajy}{{\bf{y}}}

\newcommand{\trajq}{{\bf{q}}}

\newcommand{\mds}{{\mathtt{MDS}}}
\newcommand{\swl}{{\mathtt{SwL}}}  
\newcommand{\guard}{{\mathtt{g}}}
\newcommand{\hs}{{\mathtt{HS}}}
\newcommand{\cost}{{\mathtt{cost}}}
\newcommand{\costd}{{\mathtt{cost}}}
\newcommand{\update}{{\mathtt{update}}}

\newcommand {\state} {\mathbf x}
\newcommand {\mode} {q}
\newcommand {\hybridstate} {(\state, \mode)}
\newcommand {\trace}{\vec{t}}
\newcommand {\dwellseq}{\vec{ds}}
\newcommand {\dwelltime}{dl}

\newcommand{\semantics}[1]{\ensuremath{[\![#1]\!]}}

\newcounter{myctr}

\newenvironment{myitemize}{\begin{list}{$\bullet$}
{\setlength{\topsep}{1mm}\setlength{\itemsep}{0.25mm}
\setlength{\parsep}{0.1mm}
\setlength{\itemindent}{0mm}\setlength{\partopsep}{0mm}
\setlength{\labelwidth}{15mm}
\setlength{\leftmargin}{4mm}}}{\end{list}}

\newcommand{\comment}[1]{}

%% file: intro.tex
One of the holy grails in the design of embedded and hybrid systems is to
automatically synthesize models from high-level safety and performance specifications. 
In general, automated synthesis is difficult to achieve, 
in part because design often involves human insight and intuition,
and in part because of system complexity. 
Nevertheless, in some contexts, it may be possible
for automated tools to {\em complete partial designs} 
generated by a human designer, enabling the designer
to efficiently explore the space of
design choices whilst ensuring that the 
synthesized system meets its specification.

One such problem is to synthesize the 
mode switching logic for multi-modal dynamical systems (MDS). 
An MDS is a physical system (plant) that can operate in different modes. 
The dynamics of the plant in each mode is known. 
In order to achieve safe and efficient operation, one needs to
design the controller for the plant (typically implemented in software) 
that switches between the different operating modes.  
We refer to this problem as {\em switching logic synthesis}.
Designing correct and optimal 
switching logic can be tricky and tedious for a human designer.
 
In this paper, we consider the problem of synthesizing the 
switching logic for an MDS so that the resulting system is
{\em optimal}. Optimality is formalized as minimizing a
quantitative cost measure over the long-run behavior of the system.
Specifically, we formulate cost as penalty per unit reward 
motivated by similar cost measure in Economics. 
For a given initial state, the optimal long-term behavior
corresponds to a trajectory of infinite length with infinite
number of mode switches which has minimum cost. 
So, discovering the optimal long-term
behavior requires 
\begin{myitemize}
\item discovering this infinite chain of mode switches, and 
\item the switching states from one mode to another.
\end{myitemize}
Thus, this problem would seem to involve optimization over an
infinitely-long trajectory, involving an unbounded set of parameters.
However, we reduce this problem to optimization
over bounded set of parameters representing the {\em repetitive long-term
behavior}. The key insight is that the long-term cost is essentially
the cost of the repetitive part of the behavior.
We only require the user to provide a guess of a number of switches
which could suffice to reach the repetitive behavior from an initial state. 
The system stays in repetitive behavior after reaching it
and hence, the user can pick any large enough bound. 
We consider the supersequence of all possible mode sequences with
the given number of mode switches and use the times spent in each mode
in this supersequence as the parameters for optimization. If the time
spent in a particular mode 
is zero, the mode is removed from the optimum
mode sequence. The optimization problem is then formulated as an unconstrained
numerical optimization problem which can be solved by off-the-shelf tools.
Solving this optimization problem yields the time spent in each mode
which in turn gives us the optimum mode switching sequence.
So, to summarize, for a given initial state, we obtain a sequence of 
{\em switching states} at which mode transitions must occur so as to
minimize the long-run cost.
The final step involves generalizing from a sample of switching states
to a {\em switching condition}, or {\em guard}. 
Given an assumption on the structure of guards, an inductive learning 
algorithm is used to combine switching states for different
initial states to yield the optimum switching logic for the entire
hybrid system.

To summarize, the novel contributions of this paper are as follows:
\begin{myitemize}
\item
We formalize the problem of synthesizing optimal switching logic by 
introducing the notion of long-run cost which needs to be minimized 
for optimality (Section~\ref{sec:prob}).

\item
The synthesis problem requires optimization
over infinite trajectories and not just a finite time horizon. 
We show how to reduce optimization over an 
infinite trajectory to an equivalent optimization over a bounded set of 
parameters representing the {\em limit} behavior. 
(Section~\ref{sec-formulate});

\item
We present an algorithm to solve this optimization problem for a single 
initial state 
based on unconstrained numerical optimization (Section~\ref{sec-algo}).
Our algorithm makes {\em no assumptions on the intra-mode
continuous dynamics} other than locally-Lipschitz continuity and
relies only on the ability to accurately simulate the dynamics,
making it applicable even for nonlinear dynamics;

\item
An inductive learning algorithm based on randomly sampling initial
states is used to generalize from optimal switching states 
for individual initial states to an
optimal switching guard for the set of all initial states. 
This generated switching logic is guaranteed to be the true
optimal switching logic with high probability (Section~\ref{sec-multiinit}).

\end{myitemize}
Experimental results demonstrate our approach on a range of examples
drawn from embedded systems design (Section~\ref{sec-cases}).
 
\comment{
Introduce {\em{quantitative verification}}:
from calculating 0-1 (Boolean) answers to 
calculating real numbered answers to queries about models.
Eg. calculating performance measures for designed systems.

What is a quantitative query? How do we specify a measure of
interest?  Hybrid automata that calculates the real-valued
quantities of interest.

Now, this paper is not about quantitative verification
but quantitative synthesis. 
Eg. designing optimal control, as opposed to designing
control for safety or reachability objective.

Under this larger umbrella, we consider synthesis of 
switching logic that optimizes the long-term average
of user-specified measure.~\cite{Jha10:ICCPS}
}

%% file: prob.tex
\noop{ 
Define HS = $\langle Q, X, f, d, Inv, Init\rangle$

Define HS for property = $\langle Q, X, f, d, Inv, Init, XX\rangle$
where $XX\subseteq X$ is the measures of interest.

Typically, one would be interested in the cross product.
In our application, property HS will have the same modes as system HS;
and hence, we can consider 'extension of HS with measure variables'

Define long-term average:
 given HS, HSreward with rewards $r_1,\ldots,r_l$, 
HSpenalty with penalties $p_1,\ldots,p_l$, 
 long-term average cost of a trajectory = limit as t tends to $\infty$ 
of the $\sum_{i=1}^{l} \frac{p_i}{r_i}$.

Define 
 long-term average cost of a HS for a initial state 
as the max of all trajectories starting from that initial state.

Define 
 long-term average cost of a HS 
as the max of all trajectories.

One can frame the quantitative verification problem as calculating
the long-term average cost of a given HS, HSreward, HSpenalty tuple.

Now define the synthesis variant:
HS can be viewed as MDS with a switching logic.

Given MDS, HSreward, HSpenalty, find SwL s.t. long-term average cost
of the tuple is minimal.
}

\subsection{Multimodal and Hybrid Systems}

We model a hybrid system as a combination of a
multimodal dynamical system and a switching logic.

\begin{definition}{{\bf{Multimodal Dynamical System ($\mds$).}}} 
A {\em{multimodal dynamical system}} is a tuple
$\langle Q, X, f, \init\rangle$, where
$Q := \{1,\ldots,N\}$ is a set of modes,
$X := \{x_1,\ldots,x_n\}$ is a set of continuous variables,
$f: Q\times\real^X \mapsto \real^X$
defines a vector field for each mode in $Q$,
and 
$\init \subseteq Q\times\real^X$ is a set of initial states.
The {\em{state space}} of such an MDS is
$Q\times \real^X$. 
A function 
$\traj: \real^+ \mapsto (Q\times\real^X)$
is said to be a {\em{trajectory}} of this MDS with respect to
a sequence $t_1,t_2,\ldots$ of {\em{switching times}} if
\\
(i) $\traj(0) \in \init$ and
\\
(ii) for all $i$ and for all $t$ such that $t_i < t$ and $t < t_{i+1}$, 
it is the case that
$\trajq(t) = \trajq(t_i)$ and
\begin{equation}
\frac{d{\trajx(t)}}{dt} =  f(\trajq(t), \trajx(t)),
\end{equation}
where $\trajq$ and $\trajx$ denote the projection of $\traj$ into
the mode and continuous state components.
The function $\trajx$ is continuous. The 
\em{switching sequence} is the sequence of modes
$\trajq(t_1), \trajq(t_2), \ldots$.

%
\end{definition}
If $\mds$ is a multimodal dynamical system, then its
semantics, 
denoted $\semantics{\mds}$, is the set of its trajectories 
with respect to all possible switching time sequences.
\begin{definition}{{\bf{Switching Logic ($\swl$).}}}
A {\em{switching logic}} for a multimodal system
$\mds := \langle Q,X,f,\init\rangle$ is a tuple
$\langle (\guard_{q_1q_2})_{q_1,q_2\in Q}\rangle$
where $\guard_{q_1q_2}\subseteq \real^X$ is the
{\em guard} defining the switch from mode $q_1$ to mode $q_2$.
\end{definition}

Given a multimodal system and a switching logic, we can
now define a hybrid system by considering only those
trajectories of the multimodal system that are consistent
with the switching logic.

\begin{definition}{{\bf{Hybrid System ($\hs$).}}}
A {\em{hybrid system}} is a tuple $\langle \mds,\swl\rangle$
consisting of a multimodal system 
$\mds := \langle Q,X,f,\init\rangle$
and a switching logic
$\swl := \langle (\guard_{q_1q_2})_{q_1,q_2\in Q} \rangle$.
The {\em{state space}} of the hybrid system is the same as
the state space of $\mds$. 
A function 
$\traj: \real^+ \mapsto (Q\times\real^X)$
is said to be a {\em{trajectory}} of this hybrid system
if there is a 
a sequence $t_1,t_2,\ldots$ of {\em{switching times}} such that
\\
(a)
$\traj$ is a trajectory of $\mds$ with respect to this switching
time sequence and
\\
(b) setting $t_0 = 0$, for all $t_i$ in the switching time sequence with
$i\geq 1$, 
$\trajx(t_i) \in \guard_{\trajq(t_{i-1})\trajq(t_i)}$
and for all $t$ such that $t_{i-1} < t < t_i$, 
$\trajx(t) \not\in \cup_{q\in Q}\guard_{\trajq(t_{i-1})q}$.
\end{definition}

Discrete jumps are taken as soon as they are enabled and they do not change the continuous variables. For the notion of a trajectory to be well-defined, guards are required to be closed sets.
The semantics of a hybrid system $\hs$, denoted $\semantics{\hs}$, is 
the collection of all its trajectories as defined above.

\subsection{Quantitative Measures for Hybrid Systems}
Our interest is in automatically synthesizing hybrid systems which are
{\em optimal} in the long-run. 
We define a quantitative measure on a hybrid system $\hs$ by extending 
$\hs$ with new continuous state variables.
The new continuous variables compute ``rewards'' or ``penalties''
that are accumulated over the course of a hybrid trajectory.
We also allow the new variables to be updated during discrete
transitions, which enables us to penalize or reward discrete mode
switches.

\begin{definition}{{\bf{Performance Metric.}}}
A {\em{performance metric}} for a 
given multimodal system $\mds := \langle Q,X,f,\init\rangle$
is a tuple 
$\langle \PR, f_\PR, \update\rangle$, where
$\PR := P \cup R$ is a finite set of 
continuous variables (disjoint from $X$),
partitioned into penalty variables $P$ and
reward variables $R$,
$f_\PR : Q\times\real^X \mapsto \real^\PR$ defines the vector field
that determines the evolution of the variables $\PR$,
and
$\update: Q\times Q\times\real^\PR \mapsto\real^\PR$ defines
the updates to the variables $\PR$ at mode switches.
\end{definition}

Given a trajectory 
$\traj: \real^+ \mapsto (Q\times\real^X)$ of a multimodal or hybrid system
with mode-switching time sequence $t_1,t_2,\ldots$, 
and given a performance metric, 
we define the {\em{extended trajectory}}
$\traj^e: \real^+\mapsto(Q\times\real^X\times\real^\PR)$ 
with respect to
the same mode-switching time sequence as any function 
that satisfies 
$ \traj^e(0) =  (\trajq(0),\trajx(0),\vec{0})$ and $ \traj^e(t) = (\trajq(t),\trajx(t),\trajPR(t))$,
where $\trajPR$ satisfies:
$ \frac{d\trajPR(t)}{dt}  =  f_\PR(\traj(t)) \mbox{ for all $t: t_i<t<t_{i+1}$}, $ and  $\trajPR(t_{i}) =  \update(\trajq(t_{i-1}),\trajq(t_{i}),\lim_{t\rightarrow t_{i}^-}\trajPR(t))$.

The ${\cost}$ of a trajectory 
$\traj$ is defined using its
corresponding extended trajectory 
$\traj^e$ as
\begin{equation}
\label{eqn-cost}
 \cost(\traj)  :=  \lim_{t\rightarrow\infty} \sum_{i=1}^{|P|} \frac{\trajP_i (t)}{\trajR_i (t)} 
\end{equation}
where $\trajP_i$ and $\trajR_i$ are the projection of $\traj^e$ onto the 
$i$-th penalty variable and 
$i$-th reward variable, and $|P|= |R|$.

We are only interested in trajectories where the above limit
exists and is finite. 
\noop{ 
The cost of a multimodal or hybrid system is defined as the
supremum of the costs of all the trajectories in its semantics.
\begin{equation}
 \cost(\hs) := \sup \{\cost(\traj)\mid \traj\in \semantics{\hs}\}
\end{equation}
Note that we are overloading the use of $\cost$:
it applies both to a trajectory and to a hybrid or multimodal system.
} 
We will further define $\cost$ of a part of
a trajectory from time instant $t_1$ to a time instant $t_2$
($t_2 > t_1$) as follows:
\begin{equation}
 \cost(\traj,t_1,t_2) := \sum_{i=1}^{|P|}  \frac{\trajP_i(t_2)-\trajP_i(t_1)}{\trajR_i(t_2)-\trajR_i(t_1)}
\label{eqn-costdecomposed}
\end{equation}
where $\trajP_i$ and $\trajR_i$ are components of $\trajPR$ as before.

As the definition of $\cost$  indicates,
we are interested in the {\em{long-run average}} (penalty per
unit reward) cost rather
than (penalty or reward) cost over some bounded/finite time horizon.
\forcav{Some examples of auxiliary performance variables ($\PR$) 
and $\cost$ function are described in the full version~\cite{full}.
}
\foremsoft{Some examples of auxiliary performance variables ($\PR$) 
and $\cost$ function are described below.

\begin{itemize}\itemsep=0pt
\item the {\em{number of switches}} that take place in a trajectory
 can be tracked by defining an auxiliary variable $p_1$ that has dynamics
 $\frac{dp_1}{dt} = 0$ at all points in the state space, and that
 is incremented by $1$ at every mode switch; that is,
 $$
  p_1(t_i) = \update(q,q',p_1(t_{i}^-)) = p_1(t_i^-) + 1
 $$
\item the {\em{time elapsed since start}} can be tracked by defining
 an auxiliary variable $r_1$ that has dynamics $\frac{dr_1}{dt} = 1$ at
 all points and that is left unchanged at discrete transitions; that is,
 $$
  r_1(t_i) = \update(q,q',r_1(t_{i}^-)) = r_1(t_i^-)
 $$
\item the {\em{average switchings}} (per unit time) 
 can be observed to be $\frac{p_1}{r_1}$.
 If this cost becomes unbounded as the time duration of a trajectory
 increases, then this indicates {\em{zeno}} behavior.
 Thus, if we use $p_1$ and $r_1$ as the penalty and reward variables
 in the performance metric, then we are guaranteed that
 non-zeno systems will have ``smaller'' cost and thus be ``better''.
\item
 the {\em{power consumed}} could change in different modes of a 
 multimodal system and an auxiliary (penalty) variable can track the 
 power consumed in a particular trajectory.
\item
 the {\em{distance from unsafe region}} can be tracked by an
 auxiliary reward variable that evolves based on the distance of the
 current state from the closest unsafe state.
\end{itemize}
}

\subsection{Optimal Switching Logic Synthesis}


\noop{ 
\begin{definition}{{\bf{Optimal Switching Synthesis Problem.}}}
Given a multimodal system $\mds$,
and
a performance metric, the
{\em{optimal switching logic synthesis problem}} seeks
to find a switching logic $\swl^*$ such that the cost of
the resulting hybrid system $\hs^* := \hs(\mds,\swl^*)$ 
is no more than the cost of 
an arbitrary hybrid system $\hs := \hs(\mds,\swl)$ obtained
using an arbitrary switching logic $\swl$. 
\begin{equation}
 \swl^*  := \argmin_{\swl} \{ \cost(\hs) \mid \hs := \hs(\mds,\swl) \}
\end{equation}
\end{definition}
} 

\begin{definition}{{\bf{Optimal Switching Synthesis Problem.}}}
Given a multimodal system $\mds =  \langle Q,X,f,\init\rangle$ ,
and
a performance metric, the
{\em{optimal switching logic synthesis problem}} seeks
to find a switching logic $\swl^*$ such that the cost of a trajectory
from any initial state
in the resulting hybrid system $\hs^* := \hs(\mds,\swl^*)$ 
is no more than the cost of corresponding trajectory from the same initial
state in
an arbitrary hybrid system $\hs := \hs(\mds,\swl)$ obtained
using an arbitrary switching logic $\swl$, that is,
\noop{\begin{align*}
\forall \hybridstate \in \init \; .\; \cost(\traj^*) \leq \cost(\traj) \\
where \; \traj^*(0) = \traj(0) = \hybridstate, \traj \in \semantics{\hs^*}, \traj \in \semantics{\hs}
\end{align*}
}
$
\forall \hybridstate \in \init \; .\; \cost(\traj^*) \leq \cost(\traj) \;
where \;\; \traj^*(0) = \traj(0) = \hybridstate, \traj \in \semantics{\hs^*}, \traj \in \semantics{\hs}
$
\end{definition}

We will assume, without loss of any generality, that we are given
an over-approximation of the switching logic 
$\swl^{\mathit{over}} := \langle (\guard_{qq'}^{\mathit{over}})_{q,q'\in Q}\rangle$.
In this case, the optimal synthesis problem seeks to find a 
switching logic 
$\swl^{*} := \langle (\guard^*_{qq'})_{q,q'\in Q}\rangle$
that also satisfies the constraint that
$
 \guard^*_{qq'}  \subseteq \guard^{\mathit{over}}_{qq'}  \mbox{ for all } q,q'\in Q,
$
which is also written in short as $\swl^* \subseteq \swl^{\mathit{over}}$.

The over-approximation $\swl^{\mathit{over}}$
of the switching set can be used to restrict
the search space for switching conditions.
The set $\guard^{\mathit{over}}_{qq'}$
can be an empty set if switches are disallowed from $q$ to $q'$. 
The set $\guard^{\mathit{over}}_{qq'}$
can be $\real^X$ if there is no restriction on 
switching from $q$ to $q'$.

\subsection{Running Example}

Let us consider a simple three mode thermostat controller as our running example. The multimode dynamical system describing this system is presented in Figure~\ref{fig:thermo}. The thermostat controller is described by the
tuple $\langle Q,X,f,\init \rangle$ where $Q = \{\thermoOff, \thermoHeat, \thermoCool\}$, $X= \{\thermoTemp,\thermoOut\}$, $f$ is $f_{\thermoOff}: \dot \thermoTemp = -0.1(\thermoTemp - \thermoOut)$ in mode $\thermoOff$, $f_{\thermoHeat}: \dot \thermoTemp = -0.1(\thermoTemp - \thermoOut) +0.05(80-\thermoTemp)$ in mode $\thermoHeat$ and $f_{\thermoCool}: \dot \thermoTemp = -0.1(\thermoTemp - \thermoOut) +0.15(\thermoTemp)$ in mode $\thermoCool$, and $\init = \thermoOff \times [18,20] \times [12,26]$. For simplicity, we assume that the outside temperature $\thermoOut$ does not change.
\begin{figure}[ht]
\begin{minipage}[b]{0.5\linewidth}
\begin{center}
\scalebox{0.4}{\input{thermo.pstex_t}}
\end{center}
\end{minipage}
\begin{minipage}[b]{0.5\linewidth}
\begin{align*}
& \dot{\comfort} = (\thermoTemp - 20)^2, \; \dot{\fuel} = (\thermoTemp - \thermoOut)^2\\
& \dot{\weartear} = 0, \; \dot{\thermotime} = 1 \\
& \update(\mathtt{M},\mathtt{M'},\weartear) = \weartear+ 0.5 \\
& \mathtt{for \; any \; two \; different \; modes \; M, \; M' \; in \; Q} 
\end{align*}
\end{minipage}
\caption{Thermostat Controller}
\label{fig:thermo}
\end{figure}

The performance requirement is to keep the temperature as close as possible to 
the target temperature $20$ and to consume as little 
fuel as possible in the long run. We also want to minimize the wear and tear of the heater caused by switching.
The performance metric is given by the tuple $\langle \PR, f_{\PR},\update \rangle$, where penalty variables $P = \{\comfort,$ $\fuel,\weartear\}$ denote the discomfort, fuel and wear-tear due to switching and reward variables $R = \{\thermotime\}$ denote the time spent. The evolution and update functions for the penalty and reward variables is shown in Figure~\ref{fig:thermo}. We need to synthesize the guards such that the following cost metric is minimized. Since the reward variable is the time spent, minimizing this metric means minimizing the \emph{average} discomfort, fuel cost and wear-tear of the heater. We give a higher weight ($10$) to discomfort than fuel cost and
wear-tear. 
$$
\lim_{t \rightarrow \infty} \frac { 10 \times \comfort(t) + \fuel(t) + \weartear(t) } {\thermotime(t)} 
$$

%% file: thermo.pstex_t
\begin{picture}(0,0)%
\includegraphics{thermo.pstex}%
\end{picture}%
\setlength{\unitlength}{3947sp}%
\begingroup\makeatletter\ifx\SetFigFont\undefined%
\gdef\SetFigFont#1#2#3#4#5{%
  \reset@font\fontsize{#1}{#2pt}%
  \fontfamily{#3}\fontseries{#4}\fontshape{#5}%
  \selectfont}%
\fi\endgroup%
\begin{picture}(8274,3474)(364,-3523)
\put(2101,-1036){\makebox(0,0)[lb]{\smash{{\SetFigFont{20}{24.0}{\familydefault}{\mddefault}{\updefault}{\color[rgb]{0,0,0}$+0.15(temp)$}%
}}}}
\put(4726,-661){\makebox(0,0)[lb]{\smash{{\SetFigFont{20}{24.0}{\familydefault}{\mddefault}{\updefault}{\color[rgb]{0,0,0}$\dot {temp} = -0.1(temp-out)$}%
}}}}
\put(5401,-1036){\makebox(0,0)[lb]{\smash{{\SetFigFont{20}{24.0}{\familydefault}{\mddefault}{\updefault}{\color[rgb]{0,0,0}$+0.05(80-temp)$}%
}}}}
\put(6001,-361){\makebox(0,0)[lb]{\smash{{\SetFigFont{20}{24.0}{\familydefault}{\mddefault}{\updefault}{\color[rgb]{0,0,0}HEAT}%
}}}}
\put(7051,-1936){\makebox(0,0)[lb]{\smash{{\SetFigFont{20}{24.0}{\familydefault}{\mddefault}{\updefault}{\color[rgb]{0,0,0}$g_{FH}$}%
}}}}
\put(4576,-1636){\makebox(0,0)[lb]{\smash{{\SetFigFont{20}{24.0}{\familydefault}{\mddefault}{\updefault}{\color[rgb]{0,0,0}$g_{HF}$}%
}}}}
\put(3001,-1711){\makebox(0,0)[lb]{\smash{{\SetFigFont{20}{24.0}{\familydefault}{\mddefault}{\updefault}{\color[rgb]{0,0,0}$g_{CF}$}%
}}}}
\put(376,-2086){\makebox(0,0)[lb]{\smash{{\SetFigFont{20}{24.0}{\familydefault}{\mddefault}{\updefault}{\color[rgb]{0,0,0}$g_{FC}$}%
}}}}
\put(1726,-436){\makebox(0,0)[lb]{\smash{{\SetFigFont{20}{24.0}{\familydefault}{\mddefault}{\updefault}{\color[rgb]{0,0,0}COOL}%
}}}}
\put(2476,-3061){\makebox(0,0)[lb]{\smash{{\SetFigFont{20}{24.0}{\familydefault}{\mddefault}{\updefault}{\color[rgb]{0,0,0}$\dot {temp} = -0.1(temp-out)$}%
}}}}
\put(3826,-2686){\makebox(0,0)[lb]{\smash{{\SetFigFont{20}{24.0}{\familydefault}{\mddefault}{\updefault}{\color[rgb]{0,0,0}OFF}%
}}}}
\put(526,-736){\makebox(0,0)[lb]{\smash{{\SetFigFont{20}{24.0}{\familydefault}{\mddefault}{\updefault}{\color[rgb]{0,0,0}$\dot {temp} = -0.1(temp-out)$}%
}}}}
\end{picture}%

%% file: related.tex
\section{Related Work}

There is a lot of work on synthesis of controllers for hybrid systems,
which can be broadly classified along several different dimensions.
First, based on the {\em{property of interest}}, synthesis work
broadly falls into one of two categories.
The first category finds controllers that meet some
{\em{liveness}} specifications, such as synthesizing a trajectory to drive
a hybrid system from an initial state to a desired final
state~\cite{MV99:IEEE,KooSyn:2001}, while also minimizing
some cost metric~\cite{Branicky}.
The second category finds controllers that meet some
safety specification;  
see~\cite{AsarinSyn:2000} for detailed related work in this category.
Our previous work~\cite{jha-cps10} based on combining
simulation and algorithmic learning also falls in this caregory.
Purely constraint based approaches for solving switching logic 
synthesis problem have also being used for 
reachability specifications~\cite{taly-emsoft10}.
While our work does not directly consider only safety or only
liveness requirements, both these requirements can be suitably
incorporated into the definition of ``reward'' and ``penalty''
functions that define the cost that our approach then optimizes.
While optimal control problems for hybrid systems have been
formulated where cost is defined over some finite trajectory, 
we are unaware of any work in control of hybrid
systems that attempts to formulate and solve the optimal control problem
for {\em{long-run}} costs.

The second dimension that differentiates work on controller synthesis
for hybrid systems is the space of control inputs considered; that is,
what is assumed to be controllable.
The space of controllable inputs could consist of any combination
of 
continuous control inputs, 
the mode sequence,
and the
dwell times within each mode.
A recent paper by Gonzales et al.~\cite{Tomlin10}
consider all the three control parameters,
whereas some other works either assume the
mode sequence is not controllable~\cite{Antsaklis,Caines}
or 
there are no continuous control inputs~\cite{Magnus08}.
In our work, we assume there are no continuous control inputs
and both the mode sequence and the dwell time within each
mode are controllable entities.

The third dimension for placing work on controller synthesis
of hybrid systems is the approach used for solving the 
synthesis problem.
There are direct approaches for synthesis
that compute the controlled reachable states in the style of
solving a game~\cite{AsarinSyn:2000,TomlinSyn:2000}, and abstraction-based
approaches that do the same, but on an abstraction or approximation of the
system~\cite{MoorSyn:1999,CurySyn:1998,Tab08}.  
Some of these approaches are limited
in the kinds of continuous dynamics they can handle.  They all require
some form of iterative fixpoint computation.
The other class of approaches are based on using nonlinear optimization
techniques and gradient descent~\cite{Tomlin10,Magnus08}.
Axelsson et al.~\cite{Axelsson-JOTA08} use a bi-level hierarchical optimization
algorithm with the higher level used for finding optimal mode sequence
employing single mode insertion technique, and the lower level used to find
the switching times that minimizes the cost function for fixed mode
sequence. Gonzales et al.~\cite{Tomlin10,Gonzalez-CDC10} 
extended the technique by also considering
control inputs apart from mode sequence and switching times. 
Their approach~\cite{Gonzalez-CDC10} 
can handle multiple objectives in the cost function,
can be initialized at an infeasible point and can include switching costs. 
We also use similar techniques in this paper and reduce the
controller synthesis problem to an optimization problem of a
function that is computed by performing simulations of the
dynamical system.

Notions of long-run cost similar to ours have appeared in
other areas. 
The notion of
long-run average cost is used in economics to describe 
the cost per unit output (reward) in the long-run.
In computer science,
long-run costs have been studied for graph optimization 
problems~\cite{karp-dm78}.
Long-run average objectives have also been studied for 
markov decision processes (MDPs)~\cite{markov1,markov2}.
  However, MDPs do not
have any continuous dynamics. 
Another related work is
optimal scheduling using the priced
timed automata~\cite{Rasmussen-FMSD06} in which timed
automata is extended by associating 
a fixed cost to each transition and a fixed cost rate per time unit in
a location. We consider multi-modal dynamical systems with
possibly non-linear dynamics and our cost rates
are functions of the continuous variables. Further, our interest
is in long-run cost. 
There is some recent work on
controller synthesis with budget contraints
where the budget applies in the long-run~\cite{budget}.

In contrast to existing literature,  
we present an automated synthesis algorithm to synthesize
switching logic $\swl$ for a given $\mds$ and performance metric
such that all trajectories in the hybrid system $\hs(\mds,\swl)$ 
have minimum long-term cost with respect to the given performance metric.


%% file: optimize.tex
In this section, we formulate the problem of finding switching logic
for minimum long-run cost from an initial state as an optimization problem.
Given a multimodal system $\mds = \langle Q, X, f, d, Inv, Init\rangle$, an initial state $(\mode_0, \state_0) \in Init$ and the performance metric tuple $(Y,f_Y, \update)$, we need to find the switching times $t_1, t_2, \ldots$ and the mode switching sequence $\trajq$ 
such that the corresponding trajectory $\traj$ is of minimum cost.
\small 
\begin{eqnarray}
 &&\min_{\trajq,t_1,t_2,\ldots} \cost(\traj) \;\;\; \mbox{\emph{subject to}}  \label{eqn-opt} \nonumber \\
 &&(1)[\text{Init}]: \traj(0) = (\mode_0, \state_0) \;\;
 (2)[\text{Guards}]: \forall i \; \trajx(t_i) \in \guard^{over}_{\trajq(i)\trajq(i+1)}  \nonumber \\
 &&(3)[\text{Time elapse}]: \forall t \;.\; t_i <t < t_{i+1} \; . \; \trajq(t) = \trajq(t_i), i = 1,2,\ldots;  \nonumber \\
 &&(4)[\text{Flow}]: \forall t \; \frac{d{\trajx(t)}}{dt} =  f(\trajq(t), \trajx(t))   
\end{eqnarray}
\normalsize

Since the switching sequence $t_1,t_2,\ldots$ could be of infinite
length, it is not a-priori evident how to solve the above problem.
\noop{
A naive approach to solve the above optimization algorithm requires searching
over all the mode sequences $\trajq$ and all the switching sequences 
$t_1, t_2, \ldots$ to discover the trajectory $\traj$ of minimum cost. Since trajectories are of infinite length and consequently the switching sequences
have infinite number of switching times, such a naive search
is not feasible. 
}%
In the rest of the section, we formulate an equivalent optimization problem with finite number of switching times as variables.  

Let {\em{trajectory segment}} $\traj_{[tf,te]}$ of a trajectory 
$\traj$ of length 
$L = tf-te$ be the restriction of the {\em{trajectory}} to $tf \leq t \leq te$, 
that is, $\traj_{[tf,te]}: T  \mapsto (Q\times\real^X)$ where 
$T = [tf,te] \subseteq \real^+ $
and $\traj_{[tf,te]}(t) = \traj(t)$ for $tf \leq t \leq te$.
The {\em{switching times}} of the trajectory segment is 
a finite subsequence $t_m,t_{m+1},\ldots t_n$ of the switching times
$t_1\ldots,t_m,\ldots,t_n,\ldots$
of the trajectory $\traj$ and 
$t_{m-1} < tf \leq t_m$ and $t_n \leq te < t_{n+1}$.
The special case of a trajectory segment is a {\em trajectory prefix} in which the trace starts at time $ts=0$. 

Our goal is to minimize the lifetime cost. The lifetime cost is
dominated by the the cost of the {\em limit behavior} of the system.
We are only interested in the following
stable limit behaviors (discussed in further detail in
the extended version~\cite{full}) when the lifetime cost is 
defined by the limit in Equation~\ref{eqn-cost}. 

\begin{itemize}
\item \emph{asymptotic:} for any $\epsilon$, there exists a time $t_{\epsilon}$
after which the trajectory gets asymptotically $\epsilon$-close to some state 
$(\mode_T, \state_T)$,
$|| \traj(t) - (\mode_T, \state_T) ||^2 < \epsilon$ for all $t \geq t_{\epsilon}$ where $||.,.||$ denotes the Euclidean norm, or
\item \emph{converging:} there exists a time $t_{conv}$ after which the trajectory converges, $\traj(t) = \traj(t_{conv}) $ for all $t \geq t_{conv}$, or
\item \emph{cyclic:} there exists a time $t_{cyc}$ after which the trajectory enters a cycle with period $\period$,
$\traj(t) = \traj(t+kP)$ for all $t \geq t_{cyc}$ and $k \geq 1$.
\end{itemize}

In all these cases, we can reason about the long-run cost by considering
some finite, but arbitrarily long, trajectory prefixes.
Suppose the trajectory $\traj$ is asymptotic to some hybrid state
$\traj^\inff = (\mode^\inff, \state^\inff)$.
In this case, we assume that the penalty and reward variables
$\PR$ also asymptotically approach some values
$\PR^\inff = (P^\inff, R^\inff)$.  Now consider the trajectory prefix
$\traj_{[0,te]}$. We have
\begin{eqnarray*}
&& \cost(\traj)  =  \sum_i \frac{P^\inff_i}{R^\inff_i}
\;\;\; \mbox{ and } \\
&& \cost(\traj_{[0,te]}) = \sum_i \frac{P_i(te)}{R_i(te)}
< \sum_i \frac{P^\inff_i + \epsilon}{R^\inff_i - \epsilon}
< \cost(\traj) + \delta_\epsilon
\end{eqnarray*}
Hence, by choosing $te$ appropriately, we can find a 
trajectory prefix whose cost is arbitrarily close to the
cost of the asymptotic trajectory.

\noop{
If the trajectory $\traj$ is asymptotic to some hybrid state
$\traj_T = (\mode_T, \state_T)$, then for any $\delta$, we can construct
a converging trajectory $\traj'$ with approximately the same cost
as $\traj$ in the following way.
\begin{align*}
\traj'(t) = 
\begin{cases}
\traj(t) & \forall t \leq t_{\epsilon} \\
\traj(t_{\epsilon}) & \forall t \geq t_{\epsilon}  
\end{cases}
\end{align*}
Since, the vector $f_{PR}$ field that
determines the evolution of cost and penalty is continuous,
the difference in vector field at $\traj_T$ and a hybrid state $\epsilon$ close to it is bounded by some constant $\delta_{PR}$ which can
be made arbitrarily small by suitably choosing $\epsilon$, that is,
 $$ || \traj(t_\epsilon) - \traj_T ||^2 < \epsilon \Rightarrow ||f_{PR}(\traj(t_\epsilon)) - f_{PR}(\traj_T)||^2 < \delta_{PR} $$
By choosing a $t_\epsilon$ s.t.,
$\delta_{P}/\trajP(t_{\epsilon}) \rightarrow 0$ and $\delta_{R}/\trajR(t_{\epsilon}) \rightarrow 0$,
the above converging trajectory $\traj'$ will have
$\cost(\traj') \rightarrow \cost(\traj)$. Intuitively, for asymptotic 
trajectory $\traj$, we can simulate the trajectory till the change in 
the penalty and reward variables is negligible (that is time $t_\epsilon$)
and then approximate the cost of this trajectory by the corresponding
converging trajectory $\traj'$. Thus, we only need to consider
converging and cyclic trajectories in our further discussions. 
\endnoop}

Any repetitive trajectory $\traj$ can be decomposed into a finite prefix $ \prtraj = \traj_{[0,tp]}$ followed by a trajectory segment $\reptraj = \traj_{[tp,tP]}$ repeated infinitely. We say 
$$\traj = \prtraj \; . \; (\reptraj)^\omega$$
when
$
\forall t \leq tp \;.\;\traj(t) = \prtraj(t) 
$
and
$
\forall t \geq tp \;.\;\traj(t) = \reptraj(tp+r)
$
where $r=(t-tp) \mod \period$ and $\period = tP -tp$.
The case when the trajectory converges to some hybrid state
can be treated in the same way as a repetitive trajectory.

\noop{
If the trajectory converges to some hybrid state after $t_{conv}$, $tp = t_{conv}$ and $tP$ is any time after convergence, that is, $tP > tp$. If the trajectory enters a cycle with a period $\period$ after time $t_{cyc}$, then $tp = t_{cyc}$ and $tP = tp+\period$.}
\forcav{
In Theorem~\ref{theorem-decomposecost}, we summarize how cost converges 
to a limit for trajectories with repetitive limit behavior.
The proof is presented in the full version~\cite{full}.

\begin{theorem} \label{theorem-decomposecost}
For a trajectory $\traj$ which can be decomposed into $\prtraj \; . \; (\reptraj)^\omega$, the cost of the trajectory is equal to the cost of the repetitive segment $\reptraj$, that is, 
$\cost(\traj) = \cost (\reptraj)$.
\end{theorem}
}
\foremsoft{
In Lemma~\ref{lemma-decomposecost} and 
Theorem~\ref{theorem-decomposecost}, we summarize how cost converges 
to a limit for trajectories with repetitive limit behavior.

\begin{lemma} \label{lemma-decomposecost}
For each repetition of the segment $\reptraj = \traj_{[tp,tP]}$, the change in  penalty and reward variables is constant, that is, for $P = tP - tp$.

\begin{eqnarray*}
  \forall k \geq 1 \;.\; \displaystyle P_i(tp+kP) - P_i(tp+(k-1)P)\\
 = \displaystyle P_i(tp+P) - P_i(tp)
= \displaystyle \Delta P_i \\ 
 \forall k \geq 1 \;.\;  \displaystyle R_i(tp+kP) - R_i(tp+(k-1)P) \\
= \displaystyle R_i(tp+P) - R_i(tp)
= \displaystyle \Delta R_i 
\end{eqnarray*}
\end{lemma}

\begin{proof}
The change in penalty and reward variables is given by the evolution function $f_{\PR}$ and the update on switch function $\update$. We know that $\reptraj$ is repetitive and so, $\traj(tp+kP+t) = \traj(tp+t)$ for all $t < tP-tp$ and $k \geq 1$ and hence,
$$f_{\PR}(\traj(tp+kP+t)) = f_{\PR}(\traj(tp+t))$$ 
Also, for any mode switch time $tp \leq t_i \leq tP$, $t_i' = t_i + kP$ is also a switch time because hybrid states at $t_i$ and $t_i'$ are the same. Further,
$$\update(\mode(t_{i-1}),\mode(t_i), \lim_{t\rightarrow t_{i}^-}\trajPR(t))$$
$$= \update(\mode(t_{i-1}'),\mode(t_i'), \lim_{t\rightarrow t_{i}'^-}\trajPR(t))$$
So, integrating $f_{PR}$ over continuous evolution and applying $\update$ function at mode switches, we observe that 
\begin{eqnarray*}
 \displaystyle P_i(tp+kP) - P_i(tp+(k-1)P) = \displaystyle P_i(tp+P) - P_i(tp)\\
= \displaystyle \Delta P_i \;\;\;\;\;\; \\ 
\displaystyle  R_i(tp+kP) - R_i(tp+(k-1)P) = \displaystyle R_i(tp+P) - R_i(tp)\\
 = \displaystyle \Delta R_i  
\end{eqnarray*}
\end{proof}

\begin{theorem} \label{theorem-decomposecost}
For a trajectory $\traj$ which can be decomposed into $\prtraj \; . \; (\reptraj)^\omega$, the cost of the trajectory is equal to the cost of the repetitive segment $\reptraj$, that is, 
$\cost(\traj) = \cost (\reptraj)$.
\end{theorem}
\begin{proof}
\begin{align*}
\cost(\traj)   &:=   \lim_{t\rightarrow\infty} \sum_{i=1}^{|P|} \frac{\trajP_i(t)}{\trajR_i(t)} \mathrm{\;\;[Equation~\ref{eqn-cost}]}\\
&= \lim_{t\rightarrow\infty} \sum_{i=1}^{|P|} 
\frac
{\trajP_i(tp) + \trajP_i(t) - \trajP_i(tp)}
{\trajR_i(tp) + \trajR_i(t) - \trajR_i(tp)}\\ 
&= \lim_{k\rightarrow\infty} \sum_{i=1}^{|P|} 
\frac
{\trajP_i(tp) + k \Delta \trajP_i}
{\trajR_i(tp) + k \Delta \trajR_i} \mathrm{\;\;[Lemma~\ref{lemma-decomposecost}]}\\  
&=  \frac{\Delta \trajP_i}{\Delta \trajR_i}\mathrm{\;\;[\trajP_i(}tp\mathrm{),\trajR_i(}tp\mathrm{) \;are \;finite]} \\
&= \sum_{i=1}^{|P|} \frac{\trajP_i(tP) - \trajP_i(tp)}{\trajR_i(tP) - \trajR_i(tp)}\\
&= \cost(\traj, tp, tP) \mathrm{\;\;[Equation~\ref{eqn-costdecomposed}]} \\
&= \cost(\reptraj) \mathrm{\;\;[Definition \; of \; \reptraj]} 
\end{align*}
\end{proof}
}

Using Theorem~\ref{theorem-decomposecost}, the optimization problem in 
Equation~\ref{eqn-opt} is equivalent to the following optimization problem.
Intuitively, if the repetitive part of the trajectory and the finite prefix
before the repetitive part have finite cost, then 
the long run cost of a trajectory in the limit is the cost of the 
repetitive part of the trajectory. 
More generally, to also handle the case when
the (optimal) trajectory is asymptotic, we can replace the
cyclicity requirement, $\traj(tp) = \traj(tP)$, in the
optimization problem by the weaker requirement that
the state $\traj(tP)$ at time $tP$ be very ``close'' to the
state $\traj(tp)$ at time $tp$; see also Section~\ref{sec-fixed}.

\begin{eqnarray}
 &&\min_{\trajq,t_1,t_2,\ldots} \cost(\traj) \;\;\; \mbox{\emph{subject to}}  \label{eqn-finopt} \\
\small
 &&(1)[\text{Init}]: \traj(0) = (\mode_0, \state_0) \;\;
 (2)[\text{Guards}]: \forall i \; \trajx(t_i) \in \guard^{over}_{\trajq(i)\trajq(i+1)}  \nonumber \\
 &&(3)[\text{Time elapse}]: \forall t \;.\; t_i <t < t_{i+1} \; . \; \trajq(t) = \trajq(t_i), i = 1,2,\ldots;  \nonumber \\
 &&(4)[\text{Flow}]: \forall t \; \frac{d{\trajx(t)}}{dt} =  f(\trajq(t), \trajx(t))  \nonumber \\
&& (5)[\text{Repetitive Trajectory}]:\; \traj = \prtraj \; . \; (\reptraj)^\omega \nonumber \\
&& (6)[\text{Repetitive Time}]:\; \prtraj = \traj_{[0,tp]}, \; \reptraj = \traj_{[tp,tP]} \nonumber \\ 
&& \text{ where } 0 \leq t_1 \leq \ldots t_n \leq tP \nonumber, \;  0 \leq tp < tP \nonumber
\end{eqnarray}
\normalsize

%% file: algo.tex
\noop{In this section, we present an algorithm to solve the optimization problem presented in Section~\ref{sec-formulate}. The three main challenges in to obtain the optimal trajectory are
\begin{itemize} 
\item discovering the mode switch sequence
\item finding the switching time points
\item identifying the repetitive part of the trajectory
\end{itemize} 
}
In this section, we present an algorithm to solve the above 
optimization problem. The key idea
is to construct a scalar function
$F(\trajq,t_1,t_2,\ldots,t_n, tp, tP)$ where
$\trajq$ is the switching mode sequence;
$t_1,t_2,\ldots,t_n$ are the switching times, and 
$tp, tP$ are the times denoting repetitive behavior,
such that the minimum value of $F$
is attained when the switching mode sequence and switching times
correspond to the trajectory $\traj$ with minimum long-run cost,
and $\traj_{[tp,tP]}$ is the repetitive part of the trajectory.

Once we have constructed $F$, we need to minimize $F$. 
Apart from $\trajq$, all arguments of $F$ are real-valued.
Suppose we fix $\trajq$ and 
let $F_{\trajq}(t_1,t_2,\ldots,t_n, tp, tP)$ denote the
function $F$ with fixed mode sequence $\trajq$. 
Now $F_{\trajq}$ is a function from multiple real variables
to a real, and hence
(approximate) minimization of $F$ can be performed using
\emph{unconstrained nonlinear numerical optimization}
techniques~\cite{bonnans-book06}. 
These techniques only require that
we are able to evaluate $F$ once its arguments are fixed.
This we accomplish using
numerical simulation of the multimodal
system~\footnote{We rely on simulating continuous behavior
described by ODEs in a single mode for a fixed time period and
accurate simulation of ODEs is a well-studied problem.}.

\noop{
Our algorithm for solving the constrained optimization function in
Equation~\ref{eqn-finopt} is based on defining the function $F$
and then using unconstrained nonlinear optimization techniques
to minimize $F$. We use standard techniques
for unconstrained nonlinear optimization. 
The novelty of our technique lies 
in formulating the function $F$ such that minimizing $F$
yields the solution for the optimization problem in  Equation~\ref{eqn-finopt}.
For ease of presentation, we present our
construction of $F$ in two steps. 
First, we consider the problem when the optimal model switching
sequence $\vec{q}$ is known, and later,
we show how our technique
can be used to discover $\vec{q}$ as well.  
}

\subsection{Defining $F$}\label{sec-fixed}
\noop{
Let $F_{\trajq}(t_1,t_2,\ldots,t_n, tp, tP)$ denote the
function $F$ with fixed mode sequence $\trajq$. 
Minimizing $F_{\trajq}$ over its arguments yields switching times 
$t_1,t_2,\ldots,t_n$ and $tp,tP$ such
that the trajectory $\traj$ starting from the initial state 
$(\mode_0, \state_0)$ enters repetitive 
behavior at $tp$ and the trajectory repeats with a period of 
$tP-tp$, that is, $\traj(tp) = \traj(tP)$, and 
the corresponding trajectory segment $\traj_{rep} = \traj_{[tp,tP]}$
is of minimum cost. So, minimizing
$F$ would yield the solution to the optimization problem
in Equation~\ref{eqn-finopt}. 
}


The optimization problem in Equation~\ref{eqn-finopt} is a constrained optimization problem. The constraint $ \traj = \prtraj \; . \; (\reptraj)^\omega$ requires identifying a trajectory $\traj$ starting from the given initial state $(\mode_0, \state_0)$ such that it enters repetitive behavior at time $tp$, and
$\mode(tp) = \mode(tP)$ and $\state(tp) = \state(tP)$ where $tp < tP$. We call this constraint the repetition constraint. A standard technique for solving
some constrained optimization problems is to translate it into an unconstrained optimization problem by modifying the optimization objective
such that optimization automatically enforces the constraint. 
This is done by quantifying the violation using some metric and then
minimizing the sum of the earlier minimization objective
and the weighted violation measure.
A simple
example of this approach is presented in the full version~\cite{full}.
In order to enforce the repetition constraint by suitably modifying the optimization objective, we introduce a distance function between the hybrid states. Let $d$ be the distance function between two hybrid states such that  
\begin{align*}
d((\mode_1,\state_1), (\mode_2,\state_2)) & = ||\state_1 - \state_2||^2 \;\; \text{if} \; \mode_1 = \mode_2 \;\;\; \text{and} \;\; \infty \;\; o.w.
\end{align*}
where $||\state_1 - \state_2 ||$  is the Euclidean norm. $(Q \times X, d)$ forms a metric space. So, the distance between the hybrid states is $0$ if and only if $\mode_1 = \mode_2$ and $\state_1 = \state_2$. 

Let $F(\trajq,t_1,\ldots,t_n,tp,tP) $
\begin{flalign}
&& = \begin{cases}
\cost( \traj_{[tp,tP]} ) + M \times d(\traj(tp), \traj(tP))\nonumber \\ 
\mathtt{if} \; (a)\; 0 \leq t_1 \leq \ldots t_n \leq tP, \;  tp < tP \; \text{and} \nonumber \\
\;\;\;\;\; (b)\; \forall i \; \trajx(t_i) \in \guard^{over}_{\trajq(i)\trajq(i+1)}\nonumber \\
\infty \;\;\;\;\; \mathtt{otherwise} 
\end{cases} 
\end{flalign}
where
$M$ is any positive constant and
$\traj$ is a trajectory starting from the given initial state, that is,
\begin{eqnarray*}
&&\traj(0) = (\mode_0, \state_0); \; \forall t \; t_i <t < t_{i+1} \; \trajq(t) = \trajq(t_i) \;\;,\;   i = 1,2,\ldots \\
&& \; \text{and} \; \forall t \; \frac{d{\trajx(t)}}{dt} =  f(\trajq(t), \trajx(t)) 
\end{eqnarray*}

\noop{
$F_{\trajq}$ is the restriction of the function $F$ where the
mode switching sequence is fixed to $\trajq$. 
The function $F_{\trajq}$ evaluates to a finite  value
only if the arguments satisfy conditions
$(a)$ and $(b)$. The first condition is that 
the switching times
are non-decreasing and the times for possible 
repetitive behavior satisfy $tp <tP$. The second condition is that 
the switching states $\state(t_i)$ at switching times $t_i$
lie in the user specified over-approximation of the guards,
that is, $\state(t_i) \in \guard^{over}_{\trajq(i)\trajq(i+1)}$.
If any of these two conditions are not satisfied by the
arguments, then the function $F_{\trajq}$ evaluates to infinite.
If the two conditions are satisfied, $F_{\trajq}$
is the sum of the cost of the trajectory segment $\traj_{[tp,tP]}$
and the distance between the hybrid states at time $tp$ and $tP$ weighted
with a very large constant $M$.
}
It is easy to see that
the minimum value of the function $F$ is attained when 
the hybrid states
at time $tp$ and $tP$ are the same, that is, the trajectory segment
$\traj_{[tp,tP]}$ is the repetitive part of the trajectory and
the cost of this segment is minimum. 
\noop{
With the switching states given by $\traj(t_i)$, the trajectory 
starting with the given initial state would be 
$\traj_{[0,tp]} (\traj_{[tp,tP]})^\omega$. Since, the cost
of repetitive segment $\traj_{[tp,tP]}$ is minimized, 
this trajectory is of minimum long-run cost.
}
Using Theorem~\ref{theorem-decomposecost},
we conclude that the optimization problem in 
Equation~\ref{eqn-finopt} of Section~\ref{sec-formulate}
can be reduced to the following unconstrained multivariate
numerical optimization problem
\begin{equation}
\min_{t_1,\ldots,t_n,tp,tP} \;\;\;\;\; F(t_1,\ldots,t_n,tp,tP) 
\label{eqn-distopt} 
\end{equation}

\noop{
For given values of $\langle t_1,\ldots,t_n,tp,tP \rangle$, 
$F_{\trajq}(t_1,\ldots,t_n,tp,tP)$ can be computed using a numerical
simulator. We simulate starting from initial state $ (\mode_0, \state_0)$ till 
time $tP$ with the switching times $ t_1,\ldots,t_n$ and the mode switching sequence $\trajq$. 
Starting from some initial state $ (\mode_0, \state_0)$,
we simulate the
continuous dynamics in different modes till time $tP$. We simulate 
the continuous dynamics in the first mode in the given mode sequence till time $t_1$ and then the following mode in the given sequence for time $t_2 - t_1$ time and so on. We simulate the last mode in the sequence for time $tP-t_n$.
During simulation, we also compute the reward and penalty
by recording the extended trajectory $\traj^e$ as defined in
Section~\ref{sec:prob}. The cost of the repetitive part is computed using
Equation~\ref{eqn-costdecomposed} as
$$
\cost( \traj_{[tp,tP]} ) = \cost(\traj, tp, tP) = \frac{\trajP(tP)  - \trajP(tp)}{\trajR(tP) - \trajR(tp)} 
$$   
We also record the states at times $tp$ and $tP$ to compute the distance
between them  $d(\traj(tp), \traj(tP))$.
If the two conditions required for $F_{\trajq}$ to be finite 
are satisfied, then $F_{\trajq}(t_1,\ldots,t_n,tp,tP)$ is the sum of the cost of
the repetitive part and the weighted distance.
We use a numerical nonlinear optimization engine to 
find the minimum value of the function $F_{\trajq}$. 
}
As remarked above, if the arguments of $F$ are fixed, then
$F$ can be evaluated using a numerical simulator.  Also, for
a fixed $\trajq$, we can
use a numerical nonlinear optimization engine to 
find the minimum value of the function $F_{\trajq}$. 

\subsubsection*{\underline{Running Example}}
We illustrate our technique for the running example with a fixed
sequence of modes say $\trajq = \thermoOff, \thermoHeat, \thermoOff$ starting
from the initial state $(\thermoOff,\thermoTemp = 22, \thermoOut = 16 )$. 
The outside temperature $\thermoOut$ does not change with time and remains
the same as the initial state. Only the room temperature $\thermoTemp$
changes with time.
The switching time
sequence is $t_1,t_2$. Let $tp$ denote the time when the thermostat
enters the repetitive behavior and $tP$ be the time such that
$\thermoTemp(tp) = \thermoTemp(tP)$. When $t_1 \leq t_2 \leq tp \leq tP$
and $tp < tP$, the function 
$$F_{\trajq}(t_1,t_2,tp,tP) = \cost(\traj_{[tp,tP]}) + 1000(\thermoTemp(tp) - \thermoTemp(tP))^2$$
and it is set to $2000$ otherwise (approximating infinity in the formulation
with a very high constant).
We use $ode45$ function in MATLAB~\cite{ode} 
for numerically simulating
the ordinary differential equations representing continuous dynamics
in each mode. In order to find the minimum value of $F_\trajq$ and
the corresponding arguments that minimize the function, we use the 
implementation of Nelder-Mead simplex
algorithm~\cite{nedler-compj65}.
The minimum value of $F_{\trajq}$ is obtained at
\begin{align*}
&& t_0 && t_1 && t_2 && tp && tP\\
t && 0 && 5.02 && 5.24 && 3.54 && 5.24 \\
temp && 22.0 && 19.6 && 20.2 && 20.02 && 20.2
\end{align*}

So, the switch states corresponding to the minimum long-run cost
for the given initial state  
$(\thermoOff,\thermoTemp = 22, \thermoOut = 16 )$ and given 
switching sequence of modes 
$\thermoOff, \thermoHeat, \thermoOff$
is $\guard_{HF} =\{20.2\}$ and $\guard_{FH} = \{19.6\}$.

We repeat the experiments with different initial states but 
with the same mode switching sequence. 
Even with different initial states
$(\thermoOff,\thermoTemp = 20.5, \thermoOut = 16 )$, $(\thermoOff,\thermoTemp = 21, \thermoOut = 16 )$ and $(\thermoOff,\thermoTemp = 21.5, \thermoOut = 16 )$, we obtain the same switching states in this example: $\guard_{HF} =\{20.2\}$ and $\guard_{FH} = \{19.6\}$.

When we change the mode switching sequence to $\thermoOff, \thermoHeat,$ $ \thermoOff, \thermoHeat, \thermoOff$, we discover the optimal
switching sequence to be 
\begin{align*}
&& t_0 && t_1 && t_2 && t_3 && t_4 && tp && tP\\
t && 0 && 5.02 && 5.24 && 6.73 && 6.95 && 3.54 && 6.95 \\
temp && 22.0 && 19.6 && 20.2 && 19.6 && 20.2 && 20.2 && 20.2
\end{align*}
$t_1 = 5.02, t_2 = 5.24, t_3 = 6.73, t_4 = 6.95, tp = 3.54, tP = 6.95$
which again yields the same optimal switching states 
 $\guard_{HF} =\{20.2\}$ and $\guard_{FH} = \{19.6\}$.

We observe that the optimal behavior with
respect to the given cost metric would
be to switch from $\thermoOff$ mode to $\thermoHeat$ mode at
$\thermoTemp = 19.6$ and then switch from $\thermoHeat$ to
$\thermoOff$ mode at $\thermoTemp = 20.2$ regardless of the initial
room temperature as long as the outside temperature $\thermoOut = 16$.
The optimal mode cycle is between $\thermoOff$ and $\thermoHeat$ modes. 

For an initial state with outside temperature higher than the 
outside room temperature $\thermoOut > 20$, the optimal cycle would be
between $\thermoOff$ and $\thermoCool$ modes. 
With the mode sequence $\thermoOff, \thermoCool,
\thermoOff$ and the initial state $(\thermoOff, \thermoTemp = 20.5, \thermoOut = 26 )$, we discover the optimal switching states to be $\guard_{CF} =\{20\}$ and $\guard_{FC} = \{20.3\}$.

\subsection{Finding Optimal Mode Sequence}

The algorithm above assumed that the switching mode sequence
$\trajq$ was fixed.
It can be easily adapted to also automatically discover the 
optimal switching mode sequence.
Any mode sequence starting in
mode $1$ and with atmost
 $k$ switches in a system with $N$ modes $Q = \{ 1,2,\ldots N \}$ is a 
subsequence of $1(2\;\ldots\;N\;1)^{k}$, that is, mode
$1$ followed by $(2\;\ldots\;N\;1)$ repeated $k$ times. Let dwell-time of a mode
$i$ be the time spent in the mode $t_{i+1} - t_{i}$. Given the
switching times $t_1, t_2, ... t_{Nk}$ and $tp,tP$, we define the $NZ$ function
which removes the switch times and modes from the switching sequence with
zero dwell-times, that is,
\begin{eqnarray*}
&& NZ(\overline \trajq, t_1, t_2, \ldots,t_{Nk},tp,tP)
=  (\trajq, t_{i_1}, t_{i_2}, \ldots, t_{i_K}, tp, tP)  \\
&&\text{where} \; \trajq = q_{i_1}, q_{i_2}, \ldots, q_{i_K}, 
0 < t_{i_1} < t_{i_2} < \ldots < t_{i_K} < tP\\ 
&& \text{and} \; t_m = t_{i_j} \; \text{for all} \; i_j < m < i_{j+1}
\end{eqnarray*}

For example, given the sequence of switching times $5,6,6,$ $11,12,12 $ 
and $tp = 6.5, tP = 12.5$ with the switching mode sequence 
$\overline \trajq = 1,2,3,1,2,3,1$, 
$$NZ(\overline \trajq,5,6,6,11,12,12,6.5,12.5) = (\trajq, 5,6,11,12,6.5,12.5)$$ where $\trajq = 1,2,1,2,1$.

Given a guess on the number of mode switches $k$ such that $k$ or less 
switches are needed to reach the optimal
repetitive behavior, we can use $\overline \trajq = 1(2\;\ldots\;N\;1)^k$ as the over-approximate switching
mode sequence and then find the optimal switching subsequence corresponding to the minimal long-run cost behavior using the following modified optimization formulation. 

\begin{equation}
\min_{t_1,\ldots,t_{Nk},tp,tP} \;\;\;\;\; F(NZ(\overline \trajq,t_1,\ldots,t_{Nk},tp,tP)) 
\label{eqn-finalopt}  
\end{equation}

If the optimal value returned by minimizing the above function 
is attained with the arguments $t_1^*, \ldots, t_{Nk}^*,tp^*,tP^*$,
then the optimal switching sequence $\trajq$ and the optimal switching
time sequence is given by 
$$(\trajq, t_{i_1}, \ldots, t_{i_k}, tp, tP) = NZ(\overline \trajq,t_1^*, \ldots, t_{Nk}^*,tp^*,tP^*)$$

\subsubsection*{Running Example}
We illustrate the above technique on the running example below.
Let us guess that reaching the optimal repetitive behavior from
the initial state $\thermoOff, \thermoTemp = 22, \thermoOut = 16$ takes
atmost 2 switches. We consider the mode sequence
$\thermoOff, \thermoHeat, \thermoCool, \thermoOff, \thermoHeat, \thermoCool,
\thermoOff$ which would contain all mode sequences with 2 switches
(it also contains some mode sequences with more than 2 switches).
We try to minimize the corresponding function
$F(NZ(t_1, t_2, \ldots, t_6, tp, tP))$. 

The minimum value obtained for the function $F$ with
the starting state $(\thermoOff,\thermoTemp = 22, \thermoOut = 16)$
by our optimization engine corresponds to the following
trajectory.
\begin{align*}
&& t_0 && t_1 && t_2 && t_3 && t_4 && t_5 && t_6 && tp && tP\\
t && 0 && 5.08 &&5.32 && 5.32 && 6.97 && 7.23 && 7.23 && 4.87 && 8.66 \\
temp && 22.0 && 19.6 && 20.2 && 20.2 && 19.6 && 20.2 && 20.2 && 19.7 && 19.7
\end{align*}

The optimal mode sequence and the switching times points are obtained as
\begin{eqnarray*}
\lefteqn{NZ(\overline \trajq, 5.08,5.32,5.32,6.97,7.23,7.23,4.87,8.66) = } &&
\\ &&
(\thermoOff, \thermoHeat, \thermoOff, \thermoHeat, \thermoOff, 5.08, 5.32, 6.97, 7.23, 4.87, 8.66)
\end{eqnarray*}
Since $tp = 4.87$ and $tP = 8.66$, the repetitive part of the mode sequence
is $\thermoHeat, \thermoOff$. 
The switch from mode $\thermoOff$ to $\thermoHeat$ occurs at
times $t_1$ and $t_4$. We observe that $\thermoTemp(t_1) = \thermoTemp(t_4)
= 19.6$. So, the optimal trajectory switches from $\thermoOff$ to $\thermoHeat$
at $\thermoTemp = 19.6$. The switches from $\thermoHeat$ to $\thermoCool$
and then to $\thermoOff$ occur at the same times: $t_2 = t_3$ and $t_5 = t_6$.
So, the dwell-time in the mode $\thermoCool$ is $0$ and it needs to be
removed from the optimal switching sequence. The switch into mode $\thermoOff$
occurs at times $t_3$ and $t_6$ with $\thermoTemp(t_3) = \thermoTemp(t_6)
= 20.2$. Thus, the optimal mode sequence is
$\thermoOff, (\thermoHeat, \thermoOff)^\omega$ 
and the guards discovered from this trajectory are 
$\guard_{FH} = 19.6$ and $\guard_{HF} = 20.2$.
\qed
\noop{
\begin{verbatim}
raw exp data to be removed to appendix

init = 22; outside = 16

1 2 3 1 2 3 1

fminsearch(@simthermo,[5 1 1 3 1 1 4 14])

# fminsearch(@simthermo,
[ 7.3482    0.8688    0.0737    6.4883    0.9681   
 0.2227    2.0784   16.0600])

# fminsearch(@simthermo,
[ 7.34    0.86    0.07    6.48    0.96    
0.22    2   16.5])

fminsearch(@simthermo,
[ 7.3    0.8    0.07    6.4    0.9    
0.2    2   16.5])

fminsearch(@simthermo,
[4.7    0.4    0.07    3.9    0.6    
0.2    3.1   10.2])
5.5408    0.3809    0.0899    2.7504    
0.4498    0.0923    3.9485   10.1481
fminsearch(@simthermo,
[5.5    0.4    0.08    2.75    0.4    
0.09    3.9   10.2])

fminsearch(@simthermo,
[5.4    0.33    0.0830    2.2    0.34    
0.0879    5.4   10.7])

fminsearch(@simthermo,
[4.9801    0.2269    0.0308    1.6693    
0.2607    0.0872    4.8781    8.6155])

q =

    5.0772    0.2399         0    1.6581    
0.2538    0.0896    4.8722    8.6621

timedur =

    8.6621

ans =

         0   22.0000         0         0
    4.8722   19.6860   49.0184    1.0000
    5.0772   19.6112   49.2727    1.0000
    5.3171   20.2376   53.0702    2.0000
    5.3171   20.2376   53.0702    3.0000
    6.9752   19.5901   53.7991    1.0000
    7.2290   20.2529   57.8233    2.0000
    7.2290   20.2529   57.8233    3.0000
    8.6621   19.6851   58.2287    1.0000

\end{verbatim}
}

Thus, the approach presented so far can be used to synthesize
switching conditions for minimum cost long-run
behavior for a given initial state. We need a guess on the number 
of switches $k$ such that the optimal behavior has atmost
$k$ switches. 
We summarize the guarantee of our approach for a single initial
state in the following theorem
\begin{theorem}
For a single initial state, our technique discovers the switching
states corresponding to the optimal trajectory with minimum long-run cost
if numerical optimization engine can discover global minimum
of the numerical function $F$.
\end{theorem}
The proof of the above theorem follows from the definition of
$F$. If numerical optimization engines are guaranteed
to only find local minima of $F$, our technique will
find trajectories of minimal cost. 
\noop{
Note that the function $F$ 
is discontinuous and hence, minimizing $F$ is difficult. The
adaptation to find the mode switching sequence further adds
discontinuity to the function $F$ since, eliminating modes
when the dwell time is $0$ also removes the switch cost of switching
into and out of this mode. Since $F$ is a discontinuous
function, standard numerical gradient based optimization techniques
are not very effective. 
}
We employ the Nelder-Mead simplex
algorithm as described by Lagarias et al~\cite{lagarias-jopt98,nedler-compj65} for minimizing $F$ since it is a derivative-free method and 
it can better handle discontinuities in function $F$.
We use its implementation available as the 
$fminsearch$~\cite{fmin} function in MATLAB.

%% file: multi.tex
The approach presented in Section~\ref{sec-algo} 
can find the switching
state for each mode switch along the trajectory corresponding
to optimal long-run behavior
{\em{for a given initial state}}. 
However, since systems are generally designed to operate in more than one
initial state, we need to synthesize the {\em guard condition} 
for mode switches such that the trajectory from each initial
state has optimal long-run cost. 
In this section, we present a technique for synthesizing guard conditions
in a probabilistic setting, where we assume the ability to sample initial states from
their (arbitrary) probability distribution.
Our technique samples initial states and obtains corresponding optimal switching states
for each mode switch. From the individual optimal switching states, we generalize
to obtain the guard condition for the mode switch using {\em inductive learning} (learning from examples).
In order to employ learning, we make a {\em structural assumption} on the form of guards
and use concept learning algorithms to efficiently learn
the guards from sampled switching states.\\
\\
\textbf{Structural Assumption:} 
\\
We assume that the guard condition
is a halfspace, that is, a linear inequality over the 
continuous variables $X$.\\
\\
In the rest of the section, we discuss how the existing results from 
algorithmic concept learning can be
used efficiently to learn a halfspace representing
the guard condition from the discovered
switching states for each mode-switch. 
\\
{\underline {\em Background:} } We first 
mention results which prove the efficient
learnability of halfspaces and then present an algorithm which
can be used to learn halfspaces in the probabilistically approximately correct (PAC)
learning framework~\cite{valiant-ACM84}.
In this framework, the learner receives samples marked
as positive or negative for points lying inside and outside the concept
respectively, and the goal is to 
select a generalization concept
from a certain class of possible concepts 
such that the selected concept has low 
generalization error with very high probability.
In our case,
the concept class is the set of all possible halfspaces in $\real^n$ and
the concept to be learnt is the halfspace that is the correct guard in the
optimal switching logic. 
The points in a concept to be learnt are the states in the guard and the
points outside the concept are the states outside the guard.

A halfspace
can be learnt with a very high accuracy using polynomial-sized 
sample~\cite{blumer-ACM89}.
We briefly
summarize  the relevant results from learning theory
that establish the efficient learnability of halfspaces in
the PAC learning framework. 
A concept class is said to {\em shatter} a set of points if for any
classification of the points as positive and negative, there is 
some concept in the class which would correctly classify the points.
Any concept class is associated with 
a combinatorial parameter of the class, namely,
the Vapnik-Chervonenkis (VC) dimension
defined as the 
cardinality of the largest set of points (arbitrarily labeled as positive
or negative) that the algorithm can shatter. 
For example, consider the concept class to be partitions in 
$\real^2$ using straight lines, that is, halfspaces in $\real^2$.
The straight line should separate positive points in the true concept and 
negative points outside the concept. There exist sets of 3 points that can 
indeed be shattered using this model; in fact, any 3 points that are not 
collinear can be shattered, no matter how one labels them as positive or negative. However, it can be shown 
using Radon's theorem that no set of 4 points can be 
shattered~\cite{handbook-convexgeo}. 
Thus, the VC dimension of straight lines is 3.
In general, the VC dimension for halfspaces in $\real^n$ is known to be 
$n+1$~\cite{blumer-ACM89}. The following theorem from 
Blumer et al~\cite{blumer-ACM89} establishes the
relation between efficient
learnability of a concept class in the PAC learning
framework
and VC  dimension of the concept class.

\begin{theorem} \label{thm-vc}
Let $\mathbf C$ be a concept class with a finite VC dimension $d$.
Then, any concept in $\mathbf C$ can be learnt in the following sense:
with probability at least $1-\delta$ a concept $C$ is learnt which
incorrectly labels a point with a probability of at most $\epsilon$,
where $C$ is generated using a random sample of labeled points of
size at least
$$\max (\frac{4}{\epsilon} \log \frac{2}{\delta}, \frac{8d}{\epsilon} \log \frac{13}{\epsilon} )$$. 
\end{theorem}

Since the VC dimension
for the class of halfspaces in $\real^n$ is $n+1$, 
a halfspace can be learnt in PAC learning framework 
using a sample of size at least
$$\max (\frac{4}{\epsilon} \log \frac{2}{\delta}, \frac{8n+8}{\epsilon} \log \frac{13}{\epsilon} ) $$
This, learning halfspaces in $\real^n$ 
requires samples polynomial in $n, \frac{1}{\epsilon}$ and $\frac{1}{\delta}$
and by increasing the probabilistic accuracy of the learnt halfspace 
requires polynomial increase in the number of samples. This is critical for
efficiently learning guards in our algorithm.
\\
{\underline {\em Halfspace learning algorithm:}} We 
first discuss an algorithm $\mathtt {HSinfer}$ 
which can be used to learn halfspaces
in the PAC framework from a given sample and then, describe the
switching logic synthesis algorithm.
In $\real^n$, a halfspace is given by $\theta \cdot X + \theta_0 \geq 0$ where 
$\theta \in \real^n, \theta \in \real$ and $X$ is any point in $\real^n$ which
satisfied the above inequality if and only if the point is in the concept
halfspace to be learnt. For any point $X_i$, let $Y_i$ be 
1 if $X_i$ is in the concept and $-1$ if it is outside the concept.
The algorithm below is the standard {\em Perceptron Learning} algorithm
and is known to converge after $k$ iterations where 
$k \leq 
(max_i ||X_i||)/
(\min_i \frac{Y_i(\theta X_i + \theta_0)}{|| \theta||})^2
$~\cite{fruend-ML98}.

\begin{algorithm}
\KwIn{Set of labelled sample points $\{(X_i,Y_i)\}$}
\KwOut{$\theta,\theta_0$ such that $\theta X + \theta_0 \geq 0$ is the halfspace}
Set $\theta^0 = 0, \theta_0^0 = 0, t = 0$\;
\For{each $i$}
{
        \eIf{$\theta^0 X + \theta^0_0 \geq 0$}
        {Predicted $y_i = 1$}
        {Predicted $y_i = -1$}
 
}
\While{some $i$ has $Y_i \not = y_i$}{
pick some $i$ with $Y_i \not = y_i$\;
$\theta^{t+1} := \theta^t + Y_i X_i$\;
$\theta_0^{t+1} := \theta_0^t + Y_1$\;
$t := t+1$\;
\For{each $i$}
{
        \eIf{$\theta^t X + \theta^t_0 \geq 0$}
        {Predicted $y_i = 1$}
        {Predicted $y_i = -1$}
 
}
}
return $\theta,\theta_0$
\caption{Halfspace learning algorithm $\mathtt{HSinfer}$~\cite{fruend-ML98}}
\label{alg:hs}
\end{algorithm}

{\underline {\em Learning guards:}} We now describe the 
algorithm to learn guards for multiple initial
states using the technique presented in Section~\ref{sec-algo}
and the halfspace learning algorithm $\mathtt {HSinfer}$. The algorithm
simply involves finding optimal switching points for each mode-switch
and then using halfspace learning to infer the guards. The key idea is
to use the optimal switching states as {\em positive} points for the concept learning
problem and the non-optimal states explored during optimization (which preceded the 
optimal switching states along any trajectory) as {\em negative} points since these states cannot be  
in the guard for an optimal switching logic.

\begin{algorithm}
\KwIn{$MDS(X,Q)$, initial states $I$, tolerance of generalization error $\delta$
and maximum probability of error $\epsilon$ }
\KwOut{Optimal Switching Logic $SL^{opt}$}
1. Sample initial states from $I$ for provided $\delta,\epsilon$.\\
2. For each initial state, obtain optimal trajectory in $MDS(X,Q)$ and switching states for the mode switches along the trajectory.\\
3. Label the obtained switching states as positive points.\\
4. Label the states preceding switching states along any trajectory to be 
negative points.\\
5. Using obtained sample of positive and negative states, learn the guard for the mode switch as generalization of these states using $\mathtt{HSinfer}$. \\
6. Output these guards as synthesized optimal switching logic $SL^{opt}$.
\caption{Finding optimal switching logic $SL^{opt}$ with single initial state}
\label{alg:multi}
\end{algorithm}

{\underline {\em Guarantees:}} Under the 
structural assumption that guards are halfspaces,
our PAC learning algorithm 
algorithm computes guards with probability atleast 
$1-\delta$ such that 
the probability that a guard contains
 any state which is not a switching-state 
or misses any switching-state is at most $\epsilon$. 
Further, the guards inferred by 
the above algorithm can be made probabilistically 
more and more accurate by choosing suitable values
of $\epsilon,\delta$ and considering correspondingly larger and larger samples
of initial states as given by Theorem~\ref{thm-vc}. 
For a trajectory to be a non-optimal trajectory, any one switching
point along the trajectory needs to be classified correctly.
Thus, the following theorem establishes the probabilistic guarantees
of our switching logic synthesis algorithm.
\begin{theorem}
Given a $\mds(X,Q)$, 
using random sampling
from the set of initial states which has a sample size polynomial
in $n$,$\frac{1}{\epsilon}$ and $\frac{1}{\delta}$,
Algorithm~\ref{alg:multi} synthesizes a switching logic $\swl$
with probability at least $1-\delta$ such that 
any trajectory in the synthesized hybrid system $\hs(\mds,\swl)$
is not optimal
with probability at most $m\epsilon$,
where $m$ is the number 
of guards in the switching logic, that is,
$m = |\swl| \leq  |Q|^2$ and 
$n$ is the number of variables, that is, $n = |X|$.
\end{theorem}


\subsubsection*{Running Example}
Given the set of initial states 
$16 \leq \thermoTemp \leq 26$ and $\thermoOut \in \{16,26\}$.
The set of initial states is partitioned into subsets where each
subset is a $0.1$ interval of room temperature $\thermoTemp$
and the outside temperature is $16$ or $26$. 
The guards discovered are: 
$\;\guard_{HF} : \thermoTemp \geq 20.2 \wedge \thermoOut = 16, \;\;\;\; \guard_{FH} : \thermoTemp \leq  19.6 \wedge \thermoOut = 16, \;\;\;\;
\guard_{CF} : \thermoTemp \leq 20.0 \wedge \thermoOut = 26, \;\;\;\; \guard_{FC} : \thermoTemp \geq  20.3 \wedge \thermoOut = 26
$.

%% file: exp.tex
Apart from the running example of Thermostat controller, we applied
our technique to two other case studies: (i) an Oil Pump Controller,
which is an industrial case study from~\cite{cassez-hscc09}, and
(ii) a DC-DC Buck-Boost Converter, motivated by the problem of 
minimizing voltage fluctuation 
in a distributed aircraft power system. 
We employ the implementation of Nelder-Mead simplex algorithm
 as described by Lagarias et al~\cite{lagarias-jopt98,nedler-compj65} 
 and available as the 
$fminsearch$~\cite{fmin} function in MATLAB for numerical optimization.
A more detailed discussion of experiments is available in an extended
version~\cite{full}.  

\subsection{Thermostat Controller}
If we change the cost metric in the thermostat controller 
to 
$\lim_{t \rightarrow \infty} \frac {\comfort(t) + \fuel(t) + \weartear(t) } {\thermotime(t)} $
giving equal weight to all the three penalties (instead of $10:1:1$ weight ratio
used earlier)
the optimal switching logic discovered with this cost metric are:
$\;\guard_{HF} : \thermoTemp \geq 20.0 \wedge \thermoOut = 16, \;\;\;\; \guard_{FH} : \thermoTemp \leq  18.8 \wedge \thermoOut = 16, \;\;\;\;
\guard_{CF} : \thermoTemp \leq 21.9 \wedge \thermoOut = 26, \;\;\;\; \guard_{FC} : \thermoTemp \geq  22.7 \wedge \thermoOut = 26 
$.
We observe that the room temperature oscillates closer 
to the target temperature 
when the discomfort penalty is given relatively 
higher weight in the cost metric.
This case study
illustrates that a designer can suitably define a cost metric which
reflects their priorities and, then, our technique can be used to automatically
synthesize switching logic for the given cost metric.

\subsection{Oil Pump Controller} 
Our second case study is an Oil Pump Controller, adapted from the
industrial case study in \cite{cassez-hscc09}. 
The example consists of three components - a machine which consumes
oil in a periodic manner, a reservoir containing oil, an accumulator
containing oil and a fixed amount of gas in order to put the oil under
pressure, and a pump. The simplification we make is to use a periodic
function to model the machine's oil consumption and we 
do not model any noise (stochastic variance) in oil consumption. 

The state variable is the volume $V$ of oil in the accumulator.
The system has two modes: mode $\pumpon$ when the pump is ON and mode
$\pumpoff$ when the pump is OFF. Let
the rate of consumption of oil by the machine be given by 
$m = 3*(cos(t)+1)$ where $t$ is the time. 
The rate at which oil gets filled in the accumulator is $p$. 
$p=4$ when the pump is on and $p=0$ when the pump is off.
The change in volume of
oil in the accumulator is given by the following equation
$\dot V = p-m$
where $p$ and $m$ take different values depending on the mode of operation
of the pump. For synthesis, 
we consider two different sets of requirements~\cite{cassez-hscc09}. 

In the first set of requirements, 
the volume of oil in the tank must be within some safe limit, that is,
$1 \leq V \leq 8$ and the average volume of oil in the accumulator should be
minimized. 
We model these requirements using our cost definition by 
defining one penalty variable $p_1$ and one reward variable $r_1$. 
Let the evolution of penalty $p_1$ be 
$\dot p_1 = V \; \text{if} \; 1 \leq V \leq 8, \;\; M \; \text{otherwise}$ 
where $M$ is a very large ($M \geq 10^5p_1$)
constant ($10^6$ in our experiments) and that of reward $r_1$ be $\dot r_1 = 1$.
Minimizing the cost function  
$cost1 =\lim_{t\rightarrow\infty} \frac{p_1 (t)}{r_1 (t)} $
minimizes the average volume $\lim_{t\rightarrow\infty} \frac{\int_0^t V(t)}{t}$ 
and also enforces the safety requirement $\forall t \;. \; 1 \leq V(t) \leq 8$.

In the second set of requirements, 
we add an additional requirement to those in the first set. We require that the 
the oil
volume is mostly below some threshold $V_{high} = 4.5$ in the long run.
We model this requirement by adding an additional penalty and an additional
reward variable $p_2$ and $r_2$ with evolution functions:
$\dot p_2 = 1 \; \text{if} \; V > V_{high}, \;\; 0 \; \text{otherwise}$ and
$\dot r_2 = 1 \; \text{if} \; V < V_{high}, \;\; 0 \; \text{otherwise}$.
The new cost function is $cost2 = \lim_{t\rightarrow\infty} (\frac{p_1 (t)} {r_1 (t)} + \frac{p_2 (t)}{r_2 (t)})$. 
Let $t_{high}$ be the total duration when the volume is above $V_{high}$
and $t_{low}$ be the duration that it is below $V_{high}$.
Minimizing $p_2/r_2 = t_{high}/t_{low}$ would ensure that we spend more time 
with volume
less than $V_{high}$ in the accumulator. 

The guards: $\guard_{FN}$ from $\pumpoff$ to $\pumpon$
and $\guard_{NF}$ from $\pumpon$ to $\pumpoff$ obtained for the 
above $cost1$ objective are 
$\guard_{FN}: V \leq 3.71 \;\; \guard_{NF}:V \geq 4.62$
and for $cost2$ objective are 
$\guard_{FN}: V \leq  4.07 \;\; \guard_{NF}: V \geq 4.71$.

\begin{figure}[ht]
\centering
\subfigure[Oil Pump Controller: Volume in accumulator]{
\includegraphics[width=2.2in] {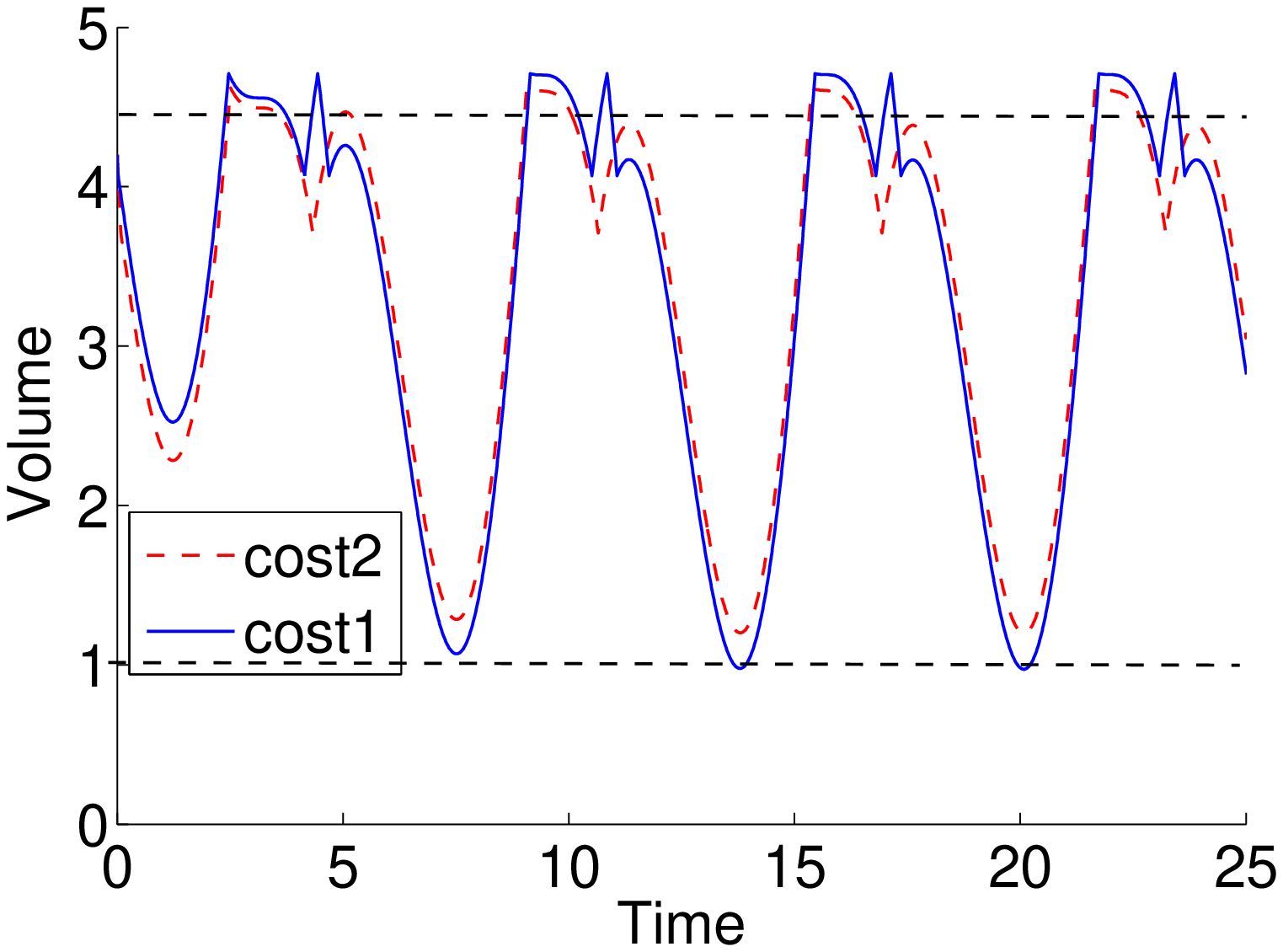}
\label{fig:oil}
}
\subfigure[DC-DC Boost Converter]{
\includegraphics[width=2.2in] {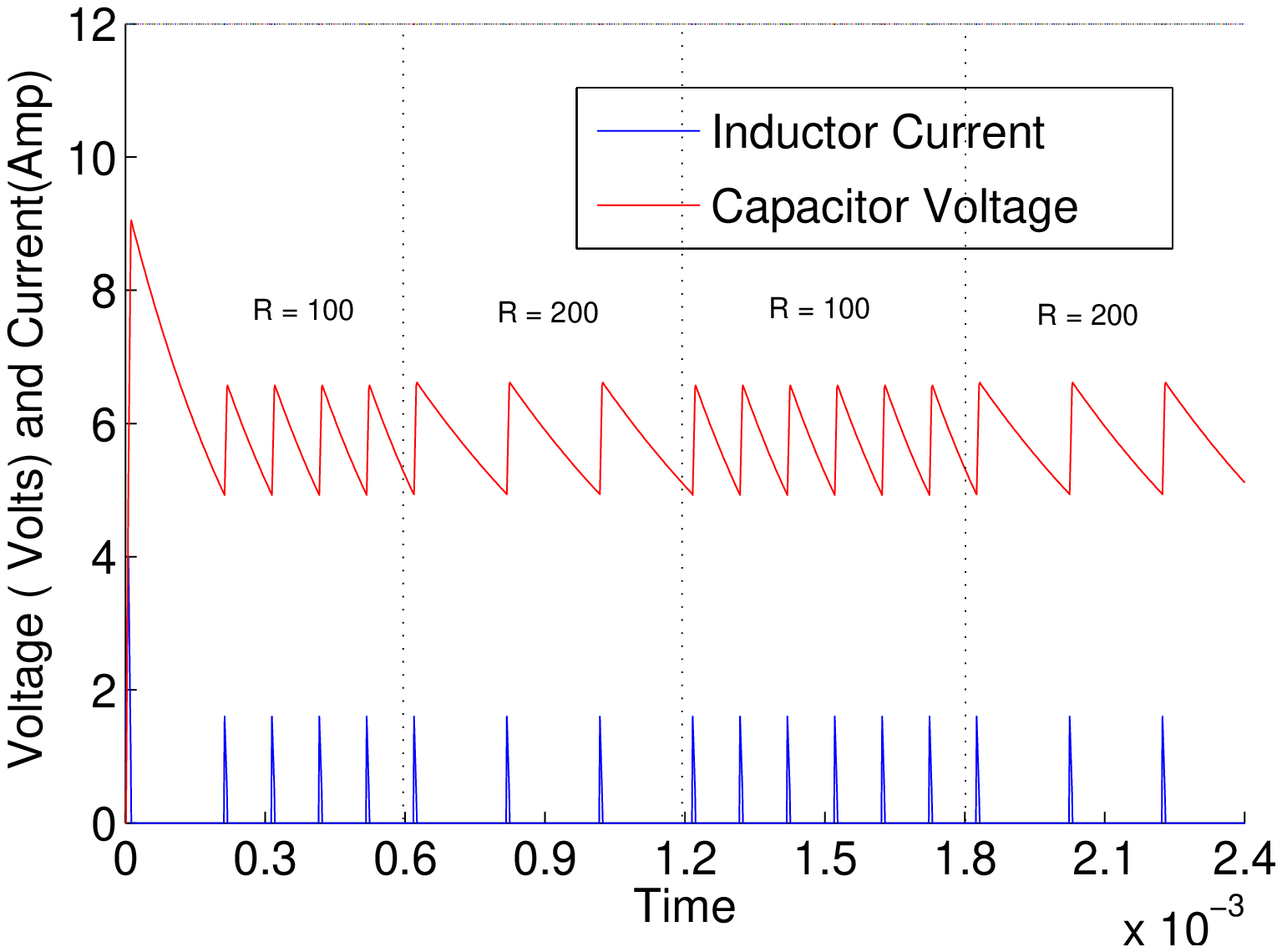}
\label{fig:boost}
}
\label{fig:subfigureExample}
\caption{Case Studies}
\end{figure}

We simulate from an initial state $V=4$ and the behavior for both objectives is
presented in Figure~\ref{fig:oil}. In both cases,
the behavior satisfies the safety property that the
volume is within $1$ and $8$. Since, we minimize oil volume, the
volume is close to the lower limit of $1$. We also observe that 
using the second
cost metric causes decrease in duration of time when oil volume 
is higher than the $4.5$ but the average volume of oil increases. 
This illustrates
how designers can use different cost metrics to better reflect their
requirements. 

\subsection{DC-DC Buck-Boost Converter}

In this case study, we synthesize switching logic for controlling
DC-DC buck-boost converter circuits described in~\cite{gupta-iscas05}.
The circuit diagram for the converter and the used parameters are presented in
the full version~\cite{full}. 
The goal is to maintain the output voltage $V_R$ across a varying 
load $R$ at some target voltage $V_d$. 
The converter can be modeled as a
hybrid system with three modes of operation.
The state space of the system is $X = \langle i_L \; u_C \rangle$
where $i_L$ is the current through the inductor and 
$u_C$ is the load voltage.
The dynamics in the three modes are given by the state space
equation $\dot X = A_k X + B_k E$ where $k =1,2,3$ is the mode
and $E$ is the input voltage.
The coefficients of the equations are
\begin{flalign*}
&A_1 = \begin{bmatrix} \frac{-rL-rs}{L} & 0\\  0 & \frac{-1}{C*(R+rC)} \end{bmatrix}, \;\;\;
B_1 = \begin{bmatrix}  \frac{1}{L} \\ 0 \end{bmatrix},\\
&A_2 = \begin{bmatrix} \frac{-rL-rd}{L} & \frac{-1}{L}&\\ \frac{R}{R+rC}*(\frac{1}{C} - \frac{rC*(rL+rd)}{L})&  \frac{-R}{R+rC}*(\frac{rC}{L}+\frac{1}{R*C})\end{bmatrix},\\
&B_2 = \begin{bmatrix} \frac{1}{L} \\ \frac{R}{(R+rC)*\frac{rC}{L}}  \end{bmatrix}, \;\;\;
A_3 = \begin{bmatrix}  0 & 0 \\ 0 & \frac{-1}{(R+rC)*C} \end{bmatrix},\;\;\; 
B_3 = \begin{bmatrix}  0 \\ 0  \end{bmatrix}&
\end{flalign*}
\noop{
The inductor acquires and stores
the energy from the source in the first mode. In the second mode,
the stored energy is transferred to the capacitor and the load.
In the third mode, the excess energy of the capacitor is
transferred to the load. 
}
We mention two key performance requirements of the DC-DC Boost Converter
described in \cite{gupta-iscas05}.
The first requirement is that the converter be resilient to load variations.
The second requirement
is to minimize the variance of the voltage across the load $V_R$ from the 
target voltage. This variance is called the ripple voltage.
We define penalty variable $p_1$ with the following evolution functions: 
$\dot p_1 = (V_R - V_d)^2$. 
We want to minimize the average deviation from the target
voltage. So, we define the reward variable $r_1$ with $\dot r_1 = 1$.
The cost function is $\cost = \lim_{t \rightarrow \infty} \frac {p_1(t)}{ r_1(t)}$.
This minimizes the average variance of $V_R$ from the target voltage
$V_d$. This corresponds to minimizing the ripple voltage. Since the load 
$R$ also
changes periodically, it also minimizes the transient variance
in voltage.

Given the dynamics in each of the three modes and the cost function,
the synthesis problem is to automatically 
synthesize the guards $\guard_{12}, \guard_{23}, \guard_{31}$ which
minimizes the cost. 
We are given the
over-approximation of the guard $\guard_{23}^{over}: i_l = 0$.
The guards obtained are as follows: $\guard_{12}: i_L > 1.9$,
$\guard_{23}:i_L = 0$ and $\guard_{23}: v_C > 4.6$.  
The system remains in the first mode until the inductor current reaches the
reference current $I_{ref}$. The system remains in the second mode until
the inductor current becomes $0$. Then, the system switches to
the third mode where it remains as long as the capacitor voltage remains over
the reference voltage $V_{ref}$. 
We simulate the synthesized
system and the behavior is shown in Figure~\ref{fig:boost}.

%% file: costEx.tex
Some examples of auxiliary performance variables ($\PR$) 
and $\cost$ function are described below.

\begin{itemize}\itemsep=0pt
\item the {\em{number of switches}} that take place in a trajectory
 can be tracked by defining an auxiliary variable $p_1$ that has dynamics
 $\frac{dp_1}{dt} = 0$ at all points in the state space, and that
 is incremented by $1$ at every mode switch; that is,
 $$
  p_1(t_i) = \update(q,q',p_1(t_{i}^-)) = p_1(t_i^-) + 1
 $$
\item the {\em{time elapsed since start}} can be tracked by defining
 an auxiliary variable $r_1$ that has dynamics $\frac{dr_1}{dt} = 1$ at
 all points and that is left unchanged at discrete transitions; that is,
 $$
  r_1(t_i) = \update(q,q',r_1(t_{i}^-)) = r_1(t_i^-)
 $$
\item the {\em{average switchings}} (per unit time) 
 can be observed to be $\frac{p_1}{r_1}$.
 If this cost becomes unbounded as the time duration of a trajectory
 increases, then this indicates {\em{zeno}} behavior.
 Thus, if we use $p_1$ and $r_1$ as the penalty and reward variables
 in the performance metric, then we are guaranteed that
 non-zeno systems will have ``smaller'' cost and thus be ``better''.
\item
 the {\em{power consumed}} could change in different modes of a 
 multimodal system and an auxiliary (penalty) variable can track the 
 power consumed in a particular trajectory.
\item
 the {\em{distance from unsafe region}} can be tracked by an
 auxiliary reward variable that evolves based on the distance of the
 current state from the closest unsafe state.
\end{itemize}

%% file: limitbehave.tex
We first discuss the possible limit behavior of a bivariate nonlinear system
using a simple example of a violin string~\cite{sastry-book}.
Using a bivariate system allows us to represent the limit behavior using
2-dimensional phase plots. 
We use linearization around the equilibrium point to analyze the limit behavior.
The limit behaviors of nonlinear systems 
and hybrid systems are more diverse than linear systems. The stable limit 
behavior can be cyclic or converging. We use MATLAB simulations
to illustrate these limit behaviors.

The violin string can be described as a second order system using the
following equation of motion.
$$M \ddot x + k x + F_b(\dot x) = 0$$

The state of the system is described using 2 variables: $x_1 = x, x_2 = \dot x$
and the evolution of the system is described as follows:
\begin{align*}
\dot x_1 &= x_2 = f_1(x_1,x_2)\\
\dot x_2 &= \frac{1}{M}(-kx_1 -F_b(x_2)) = f_2(x_1,x_2)
\end{align*}
where $F_b(x_2) = -((x_2 - b) + c)^2 - d$ and the constant values are chosen as
$$M=3,k=3,b=1,c=2,d=3$$

At equilibrium point $(x_1^*, x_2^*)$, the left-hand side of the evolution
should be zero.

So,
\begin{align*}
\dot x_1^* = x_2^* = 0\\
\dot x_2^* =  \frac{1}{M}(-kx_1^* -F_b(x_2^*)) = 0
\end{align*}

Using the given values, $x_1^* = \frac{4}{3}, x_2^* = 0$.

Jacobean linearization around $(x_1^*, x_2^*)$ can be used to analyze 
the nature of
the limit behavior at equilibrium.

$$Df = \begin{bmatrix} \partial f_1 / \partial x_1 & \partial f_1 / \partial x_2
\\ \partial f_2 / \partial x_1 & \partial f_2 / \partial x_2 \end{bmatrix}$$

To analyze the limit behavior, $$|Df |_{x^*} - \lambda I | = 0$$,
the nature of the eigen-values ($\lambda$) can be used to accurately predict the limit
behavior for {\em linear systems}. The following cases are possible
for a linear system:
\begin{itemize}
\item If both eigen-values are real and negative, the equilibrium is
a {\em STABLE NODE}, that is, the system converges to the equilibrium.
\item If both eigen-values are real and positive, the equilibrium is 
a {\em UNSTABLE NODE}, that is, the system will diverge.
\item If real part of one eigen-value is positive and another is negative, 
the equilibrium
is a {\em SADDLE}, that is, it diverges from all initial states except
a line.
\item If the eigen-values are complex and both real parts are positive,
the equilibrium is a {\em STABLE FOCUS}, that is the system
converges to the equilibrium going through a damping cycle. 
\item If the eigen-values are complex and both real parts are negative,
the equilibrium is a {\em UNSTABLE FOCUS}, that is the system
diverges through increased oscillations. 
\item If the eigen-values are purely imaginary, then the equilibrium is a
{\em CENTER} and the system enters a cycle.
\end{itemize}

Thus, the {\em stable} limit behavior for a linear system is either
converging or cycling.
For a non-linear system, linearization only provides
a hint regarding the limit behavior. It is possible to enter limit cycles 
for a non-linear system even if the linearization predicts an 
{\em UNSTABLE FOCUS}. Analytical techniques are not sufficient to predict 
limit behaviors of nonlinear and hybrid systems. Simulation is used 
to find the limit behaviors. But all {\em STABLE} behaviors are always
either converging or cyclic.

In our example of voilin strings, the eigen values for the given parameters
are
$$\lambda = \frac{1}{3} \pm \frac{2\sqrt{2}}{3}j$$

Thus, linearization predicts that the system would enter an unstable focus but
simulation shows the the system enters a limit cycle from any initial state
other than $(\frac{4}{3},0)$. If the system starts from $(\frac{4}{3},0)$,
it stays at the initial state. This illustrates the phenomenon of
{\em BIFURCATION}, that is, different initial states show different limit
behaviors in non-linear and hybrid systems.

SIMULATION

possibly with changed value of b to show center, stable focus as well apart from
limit cycle.

%% file: toyopt.tex
We recall a standard procedure to solve
constrained optimization problem with a small example. This technique
is used to solve the optimization problem with the Equation~\ref{eqn-finopt}.
Consider the following example,
\begin{eqnarray}
\min_{x,y} && \frac{1}{x^2+y^2}  \nonumber \\ \label{eqn-toy} 
\mbox{\emph{subject to}} && x + y = 10  \nonumber
\end{eqnarray}
The above optimization problem can be transformed into the following equivalent
optimization problem without constraints by suitably modifying the optimization
objective.
\begin{eqnarray}
\min_{x,y} \; \frac{1}{x^2+y^2} + M (x+y-10)^2  \nonumber \\ \label{eqn-toy1} 
 \mbox{where}\; M \; \mbox{is any positive number,} \nonumber \\ 
 \mathtt{say\; } M = 100  \nonumber 
\end{eqnarray}

%% file: lemmadecompose.tex
\begin{proof}
The change in penalty and reward variables is given by the evolution function $f_{\PR}$ and the update on switch function $\update$. We know that $\reptraj$ is repetitive and so, $\traj(tp+kP+t) = \traj(tp+t)$ for all $t < tP-tp$ and $k \geq 1$ and hence,
$$f_{\PR}(\traj(tp+kP+t)) = f_{\PR}(\traj(tp+t))$$ 
Also, for any mode switch time $tp \leq t_i \leq tP$, $t_i' = t_i + kP$ is also a switch time because hybrid states at $t_i$ and $t_i'$ are the same. Further,
$$\update(\mode(t_{i-1}),\mode(t_i), \lim_{t\rightarrow t_{i}^-}\trajPR(t))$$
$$= \update(\mode(t_{i-1}'),\mode(t_i'), \lim_{t\rightarrow t_{i}'^-}\trajPR(t))$$
So, integrating $f_{PR}$ over continuous evolution and applying $\update$ function at mode switches, we observe that 
\begin{eqnarray*}
 \displaystyle P_i(tp+kP) - P_i(tp+(k-1)P) = \displaystyle P_i(tp+P) - P_i(tp)\\
= \displaystyle \Delta P_i \;\;\;\;\;\; \\ 
\displaystyle  R_i(tp+kP) - R_i(tp+(k-1)P) = \displaystyle R_i(tp+P) - R_i(tp)\\
 = \displaystyle \Delta R_i  \;\;\;\;\;\;\\
\end{eqnarray*}
\end{proof}

%% file: multiproof.tex
\begin{proof}
There can be 
three possible scenarios for optimal trajectories from two
different initial states illustrated in
Figure~\ref{fig:noninter}.
\begin{figure}[htpb]
\begin{center}
\scalebox{0.5}{\input{noninter.pstex_t}}
\caption{Optimal trajectories from two initial states }
\label{fig:noninter}
\end{center}
\end{figure}
\begin{itemize}
\item One of the initial state is reachable from another
along the optimal trajectory of the later. 
\item There is some state $(\mode',\state')$ reachable from
both initial states along their optimal trajectories.
\item There is no state $(\mode',\state')$ reachable from
both initial states along their optimal trajectories.
\end{itemize}
In the first two cases, since the optimal trajectory must 
be deterministic, it follows that the trajectories after 
their common state are same and hence, the trajectories
have common repetitive part. In the third case,
there is no state reachable in  both optimal trajectories. 
\end{proof}